%
\input phyzzx
\catcode`\@=11 
\def\papers{\papersize\headline=\paperheadline
\footline=\paperfootline}
\def\papersize{\hsize=40pc \vsize=53pc \hoffset=0pc \voffset=-2pc
   \advance\hoffset by\HOFFSET \advance\voffset by\VOFFSET
   \pagebottomfiller=0pc
   \skip\footins=\bigskipamount \normalspace }
\catcode`\@=12 
\papers
%
%

\Pubnum={NEIP-99-020 \cr
{\tt hep-th/9912150} \cr
}

\titlepage
\title{{\bf M-theory on $S^1/{\bf Z}_2$: new facts from a careful analysis}
\foot{Work partially supported by the
Swiss National Science Foundation.}
 }
\author{Adel~Bilal, Jean-Pierre Derendinger and Roger Sauser
}
\vskip .5cm
\address{
Institute of Physics, University of Neuch\^atel, 
CH-2000 Neuch\^atel, Switzerland \break
{\tt adel.bilal,\ jean-pierre.derendinger,
\ roger.sauser @iph.unine.ch }
}

\vskip 0.5cm
\abstract{\noindent
We carefully re-examine the issues of solving the modified 
Bianchi identity, anomaly cancellations and flux quantization 
in the $S^1/{\bf Z}_2$ orbifold of M-theory using the 
boundary-free ``upstairs" formalism, avoiding several 
misconceptions present in earlier literature. 
While the solution for 
the four-form $G$ to the modified Bianchi identity appears to
depend on an arbitrary parameter $b$, we show that requiring $G$ 
to be globally well-defined, i.e. invariant under small and 
large gauge and local Lorentz transformations, fixes $b=1$. 
This value also is necessary for a consistent reduction to the
heterotic string in the small-radius limit. Insisting on properly
defining all fields on the circle, we find that 
there is a previously unnoticed additional contribution to the 
anomaly inflow from the eleven-dimensional topological term. 
Anomaly cancellation then requires a {\it quadratic} relation 
between $b$ and the combination $\lambda^6/\kappa^4$ of the gauge
and gravitational coupling constants $\lambda$ and $\kappa$.
This  contrasts with previous beliefs that anomaly cancellation
would give a cubic equation for $b$. We observe that our 
solution for $G$ automatically satisfies integer or half-integer
flux quantization for the appropriate cycles.
We explicitly write out the anomaly cancelling terms of 
the heterotic string as inherited from the M-theory approach.
They differ from the usual ones by the addition of a 
well-defined local counterterm. We also show how five-branes enter 
our analysis.
}

%

\endpage
\pagenumber=1

\def\PL #1 #2 #3 {Phys.~Lett.~{\bf #1} (#2) #3}
\def\NP #1 #2 #3 {Nucl.~Phys.~{\bf #1} (#2) #3}
\def\PR #1 #2 #3 {Phys.~Rev.~{\bf #1} (#2) #3}
\def\PRL #1 #2 #3 {Phys.~Rev.~Lett.~{\bf #1} (#2) #3}
\def\CMP #1 #2 #3 {Comm.~Math.~Phys.~{\bf #1} (#2) #3}
\def\d{\delta}
\def\D{\Delta}

\def\b{\beta}
\def\o{\omega}
\def\O{\Omega}
\def\m{\mu}
\def\n{\nu}
\def\r{\rho}
\def\l{\lambda}
\def\s{\sigma}
\def\k{\kappa}
\def\g{\gamma}
\def\f{\phi}

\def\tr{{\rm tr}\, }
\def\Tr{{\rm Tr}\, }
\def\til{\widetilde}
\def\wh{\widehat}
\def\e{\epsilon}

\def\to{\rightarrow}
\def\C{{\cal C}}
\def\orb{S^1/{\bf Z}_2}
\def\xe{x^{11}}
\def\ii{{\til I_{4,i}}\, }
\def\ij{{\til I_{4,j}}\, }
\def\ik{{\til I_{4,k}}\, }
\def\iu{{\til I_{4,1}}\, }
\def\id{{\til I_{4,2}}\, }
\def\oi{{\til\o_i}}
\def\oiu{{\til\o_i^1}}

\REF\HWI{P. Ho\v rava and E. Witten, {\it Heterotic and type I 
string dynamics from eleven
dimensions}, \NP B460 1996 506 , {\tt hep-th/9510209}.}

\REF\HWII{P. Ho\v rava and E. Witten, {\it Eleven-dimensional 
supergravity on a manifold with
boundary}, \NP B475 1996 94 , {\tt hep-th/9603142}.}

\REF\DEALWIS{S.P. de Alwis, {\it Anomaly cancellation in 
M-theory}, 
\PL B392 1997 332 , {\tt hep-th/9609211}.}

\REF\VAFAWITTEN{C. Vafa and E. Witten, {\it A one-loop test of 
string duality}, \NP B447 1995 261 ,
{\tt hep-th/9505053}.}

\REF\DUFF{M.J. Duff, J.T. Liu and R. Minasian,  {\it Eleven 
dimensional origin of
string/string duality: a one loop test}, \NP B452 1995 261 , 
{\tt hep-th/9506126}.}

\REF\WITTENFIVE{E. Witten, {\it Five-branes and M-theory on 
an orbifold}, \NP B463 1996 383 , {\tt hep-th/9512219};
{\it Five-brane effective action in M-theory}, 
J. Geom. Phys. {\bf 22} (1997) 103, {\tt hep-th/9610234}.}

\REF\CJS{E. Cremmer, B. Julia and J. Scherk, {\it 
Supergravity theory in 11 dimensions},
Phys. Lett.  {\bf 76B} (1978) 409.}

\REF\DM{E. Dudas and J. Mourad, {\it On the strongly 
coupled heterotic string}, 
\PL B400 1997 71 , {\tt hep-th/9701048}.}

\REF\CONRAD{J.O. Conrad, {\it Brane tensions and coupling 
constants from within M-theory}, Phys. Lett. {\bf B241} 
(1998) 119, {\tt hep-th/9708031}.}

\REF\LU{J.X. Lu, {\it Remarks on M-theory coupling constants 
and M-brane
tension quantizations},  {\tt hep-th/9711014}.}

\REF\FAUXI{M. Faux, {\it New consistent limits of M-theory}, 
{\tt hep-th/9801204};
{\it Confluences of anomaly freedom 
requirements in M-theory}, {\tt hep-th/9803254}.}

\REF\HARMARK{T. Harmark, {\it Coupling constants and 
brane tensions from anomaly cancellation in M-theory}, 
\PL B431 1998 295 , {\tt hep-th/9802190}.}

\REF\FLO{M. Faux, D. L\"ust and B.A. Ovrut, {\it Intersecting 
orbifold planes and local
anomaly cancellation in M-theory}, \NP B554 1999 437 , 
{\tt hep-th/9903028}.}

\REF\WITTENFLUX{E. Witten, {\it On flux quantization in 
M-theory and the effective action},
J. Geom. Phys. {\bf 22} (1997) 1, {\tt hep-th/9609122}.} 

\REF\WITTENSTRONG{E. Witten, {\it Strong coupling 
expansion of Calabi-Yau compactification}, Nucl. Phys. 
{\bf B471} (1996) 135, {\tt hep-th/9602070}.}

\REF\BDS{A. Bilal, J.-P. Derendinger and R. Sauser, work 
in progress.}

\REF\AGDPM{L. Alvarez-Gaum\'e, S. Della Pietra and G. Moore, 
{\it Anomalies and odd
dimensions}, Ann. Phys. {\bf 163} (1985) 288.}

\REF\ROHMWITTEN{R. Rohm and E. Witten, {\it The 
antisymmetric tensor field 
in superstring theory}, Ann. Phys. {\bf 170} (1986) 454.}

\REF\AGW{L. Alvarez-Gaum\'e and E. Witten, {\it Gravitational 
anomalies}, \NP B234 1984 269. \hfil\break
L. Alvarez-Gaum\'e and P. Ginsparg, {\it The structure of 
gauge and gravitational anomalies}, Ann. Phys. 
{\bf 161} (1985) 423, erratum-ibid {\bf 171} (1986) 233.} 

\REF\GSW{M. B. Green, J. H. Schwarz and E. Witten, 
{\it Superstring theory}, Cambridge University Press, 
Cambridge, 1987,  vol. 2. }

\REF\LOW{A. Lukas, B. Ovrut and D Waldram, {\it The ten-dimensional 
effective action of strongly coupled heterotic string theory}, Nucl.
Phys. {\bf B540} (1999) 230, {\tt hep-th/9801087}.}

\REF\NM{D.~Freed, J.~A.~Harvey, R.~Minasian and G.~Moore, 
{\it Gravitational anomaly cancellation for M-theory fivebranes},
Adv. Theor. Math. Phys.  {\bf 2}  (1998) 601,
{\tt hep-th/9803205}.}

\REF\LOWCY{A. Lukas, B. Ovrut and D. Waldram, {\it Non-standard 
embedding and five-branes in heterotic M-Theory},  Phys. Rev. 
{\bf D59} (1999) 106005, {\tt hep-th/9808101}.}

{\bf \chapter{Introduction and summary}}

\noindent
Ever since the emergence of M-theory and its relation to 
the various perturbative string theories, considerations 
of anomaly cancellation have played a crucial role. 
In particular, in reconstructing the strongly-coupled 
$E_8\times E_8$ heterotic 
string from M-theory compactified on $\orb$ [\HWI, \HWII], an 
important ingredient was the observation that the gauge and 
gravitational anomaly polynomial of the $E_8\times E_8$
heterotic string can be written as a sum of two terms, 
each being associated with one $E_8$ factor, and that each 
of these terms separately factorises as required by anomaly 
cancellation through a Green-Schwarz mechanism. This 
enables local anomaly cancellation on each of the two 
ten-dimensional $\orb$ fixed planes in the M-theory 
picture [\HWI, \HWII, \DEALWIS]. The anomaly-cancelling terms in 
M-theory are of two types: one is a Green-Schwarz term 
$\int G\wedge X_7$ where $G$ is the (modified) field 
strength of the three-form $C$ and $X_7$ is a purely gravitational
Chern-Simons type seven-form. The presence of this Green-Schwarz 
term was well-known from string duality [\VAFAWITTEN, \DUFF] and
as a cancelling term for the M-theory five-brane anomaly 
[\DUFF, \WITTENFIVE]. The second contribution is anomaly 
inflow  from the topological interaction term of 
eleven-dimensional supergravity [\CJS] 
$\int C\wedge G\wedge G$. In 
uncompactified eleven-dimensional M-theory both terms are 
gauge and local Lorentz invariant and no anomaly needs 
to be cancelled in this odd dimension. However,
orbifold compactifications like $\orb$ involve chiral 
projections on the even-dimensional fixed planes and 
then there are chiral anomalies which need to be 
cancelled. On the other hand, it has been shown in [\HWII] 
that closure of the supersymmetry algebra for the $\orb$ 
orbifold implies a modification of the Bianchi identity 
$d G=0$, and this is why the $\int G\wedge X_7$ and $\int 
C\wedge G\wedge G$ terms have non-vanishing anomalous
transformations under gauge and local $SO(9,1)$ Lorentz 
transformations on the fixed planes. This is the basic 
mechanism at the origin of local anomaly cancellation.

In ref. [\HWII], the modified Bianchi identity $dG\ne 0$ 
was solved in a particular way and it was concluded that 
anomaly cancellation requires a certain fixed ratio 
$\l^3/\k^2$ of the gauge coupling $\l$ and the gravitational
coupling $\k$. Subsequent analyses [\DM--\FLO] 
have in particular emphasized
that there actually is a one-parameter family of 
solutions $G(b), \ b\in {\bf R}$, to the modified Bianchi 
identity.\foot
{Of course, to any solution $G^{(0)}=d C + \ldots\,$ of 
$d G=\ldots$ one can add a $d A^{(3)}$ with any three-form 
$A^{(3)}$. The point is that most of these $A^{(3)}$ 
can be reabsorbed into $C$. So the only relevant $A^{(3)}$ 
must be made from the gauge and Lorentz Chern-Simons three-form 
$\o^{(3)}$ on the fixed planes so that essentially the only 
freedom is $A^{(3)}\sim b \o^{(3)}$ with one real parameter 
$b$.}
It was concluded that anomaly cancellation alone does not 
fix the ratio $\l^3/\k^2$ but relates it to a {\it cubic} 
polynomial in this parameter $b$. It was argued 
that this parameter and hence $\l^3/\k^2$ 
can be fixed if one also takes into account the quantization 
of the flux of $G$ [\WITTENFLUX].

In this paper, we carefully reanalyse the issues of 
solving the modified Bianchi identity, anomaly 
cancellation and flux quantization for the $\orb$ orbifold. 
We use the ``upstairs" 
formalism where one works on the {\it boundary-free} 
circle $S^1$ and imposes a ${\bf Z}_2$ projection on the
fields. However, contrary to most of the previous papers,
we insist on defining properly all fields on 
$S^1$, i.e. fields should be periodic. This immediately 
rules out the 
use of the step function $\theta(\xe)$ as a primitive 
of a delta function $\d(\xe)$ on the circle. 
Using instead a correctly periodically defined function 
$\e(\xe)$ such that $\e'\sim \d-{1\over 2\pi}$ turns out 
to be crucial to obtain a consistent solution $G$ to the 
modified Bianchi identity. We carefully investigate gauge 
and local Lorentz invariance of $G$. It turns out that 
insisting on invariance under {\it large} transformations, 
i.e. insisting on having a globally well-defined $G$, is 
very powerful. It not only implies the well-known cohomology 
condition [\WITTENSTRONG], but it also fixes the parameter 
$b$ of the solution $G$ to be 1, as long as there are 
topologically non-trivial gauge or gravity configurations 
with $\int (\tr F_i^2 - {1\over 2} \tr R^2)\ne 0$. The same 
conclusion $b=1$ is obtained if five-branes are present. The 
value $b=1$ also appears to be the only one which allows a safe 
truncation to the perturbative heterotic string: although 
the zero-modes on the circle correctly give the desired 
relation $H=dB -\o_{\rm YM}+\o_{\rm L}$ for all values of 
$b$, the neglected higher modes are gauge and local Lorentz 
invariant only if $b=1$. Note that these arguments leading 
to $b=1$ are {\it not} 
a consequence of $G$-flux quantization. We note that the 
downstairs approach automatically incorporates a 
correspondingly fixed value of $b$.

Looking carefully at anomaly cancellations also yields 
some surprises. We do the analysis for arbitrary parameter 
$b$. Again the use of the periodic $\e(\xe)$ is crucial, 
in particular the constant term in $\e'\sim\d-{1\over 2\pi}$ 
plays an important role. Due to this term, the relation 
between $b$ and the ratio $\l^3/\k^2$ required by anomaly 
cancellation is drastically modified with respect 
to previous analyses [\HWII, \DM--\FLO]: the terms cubic in $b$ 
cancel, and one is left with a 
{\it quadratic} equation for $b$ namely $b^2= 
{12\over (4\pi)^5} \l^6/\k^4$.
Also, to obtain this equation, one has to evaluate products 
of the form $\d(\xe) \e(\xe)\e(\xe)$ where $\e(\xe)$ is our 
periodic generalisation of the step function. But of course, 
this product is ill-defined and should be regularised. Any 
reasonable regularisation yields $\d(\xe) \e(\xe)\e(\xe)\to 
{1\over 3}\, \d(\xe)$ providing an extra crucial factor 
of ${1\over 3}$ in the above equation determining the 
parameter $b$ from anomaly cancellations in terms of 
$\l^3/\k^2$. This completely modified, now quadratic 
equation will have important consequences when considering 
further orbifold compactification [\FLO], e.g. on $T^4/{\bf Z}_2$, 
where the analysis has encountered certain difficulties.\foot
{
We are grateful to D. L\"ust for sharing his insights on 
this point.
} 
This will be examined elsewhere [\BDS].  We also show how the
discussion of anomalies is affected by the presence of 
five-branes. 

We then conclude that $b$ must be one, and hence 
$\l^6 = {(4\pi)^5\over 12} \k^4$. Indeed, in any 
topologically non-trivial sector of the theory where 
$\int (\tr F_i^2 - {1\over 2} \tr R^2)\ne 0$ we have 
$b=1$. Anomaly cancellation then fixes the ratio 
$\l^6/\k^4= {(4\pi)^5\over 12} b^2= {(4\pi)^5\over 12}$. 
But this ratio does not depend on whether the integral is 
zero or not, so that by this same equation $b=1$ always 
(strictly speaking: $b^2=1$).
It is worthwhile noticing that $G$ differs from $d C$ also 
away from the fixed planes since $b\ne 0$.  
We further note that the Green-Schwarz 
term $\int G\wedge X_7$ also automatically 
ensures cancellation of the five-brane anomalies without 
any need of further modification. As a consistency check 
of our solution  we show how the 
combination of the topological term and the Green-Schwarz 
term in eleven dimensions 
leads to the Green-Schwarz term of the heterotic string.

Another subtle point is flux quantization of $G$. In the 
``downstairs" approach where one considers M-theory 
compactified on the interval $I=\orb$, i.e. in the 
presence of ten-dimensional boundary planes of the 
eleven-dimensional space-time, there 
is no modification of the Bianchi identity $d G=0$, but 
there is a non-trivial boundary condition on $G$ [\HWII] 
enabling the necessary anomaly inflow to cancel 
the one-loop anomaly on the boundary planes. This boundary 
condition does not admit a free parameter
like $b$ (i.e. it is equivalent to a fixed value of $b$) 
and reads $G\vert_{\rm boundary} = {\k^2\over \l^2} 
(\tr F_i^2-{1\over 2} \tr R^2 )\vert_{\rm boundary}$. 
Witten concluded [\WITTENFLUX] that for  any four-cycle 
$\C_4$ the flux ${1\over (4\pi)^2} 
{\l^2\over \k^2} \int_{\C_4} G$ 
has to be integer or half-integer. In particular, this 
ensures that the membrane functional integral is well-defined. 
Naively the latter would seem to be 
well-defined only for integer flux, but it was shown in 
ref. [\WITTENFLUX] that in the case of half-integer 
flux the three-dimensional membrane functional 
integral has a parity 
anomaly [\AGDPM] which precisely cancels the sign ambiguity 
due to the half-integrality of the $G$-flux.
In the boundary-free upstairs approach where the Bianchi 
identity for $G$ is modified this same flux quantization 
should appear as a consequence of our solution. We will show 
that this is indeed the case.
Any four-cycle $\C_4$ can be written as a sum of four-cycles 
not involving the $\xe$-direction and of four-cycles of the 
form 
$S^1\times \C_3$ with $\C_3$ not in the $\xe$-direction. In the 
first case our solution $G$ to the modified Bianchi identity 
straightforwardly yields Witten's 
result of integer or half-integer flux (provided one 
correctly relates $\k^2_{\rm upstairs}$ 
and $\k^2_{\rm downstairs}$ with the necessary factor of 
$1/2$ [\CONRAD]). 
Four-cycles of the second type wrap around the circle $S^1$. 
There is no analogue of these cycles in the downstairs 
approach, and we cannot conclude on any quantization from 
the results of [\WITTENFLUX]. Indeed, for such cycles, 
we find that the $b$ 
dependence trivially cancels (as always when integrating over 
the circle, thanks to the modification of the 
step function in order to be  periodic!) and one 
is left with an integral 
$A\equiv \int_{S^1\times \C_3} d C 
- {1\over (4\pi)^2} {\l^2\over \k^2} 
\sum_i \int_{\C_3} \til\o_i$
where the sum is over the two fixed planes and $\til\o_i$ 
is a combination of gauge and Lorentz Chern-Simons 
three-forms on 
the $i^{\rm th}$ plane. This integral is closely 
related to the flux $\int H = \int(dB-\o_{\rm YM}
+\o_{\rm L})$ in the 
heterotic string. The latter was discussed some time ago 
[\ROHMWITTEN] and it was concluded (when appropriately 
normalised) to be of the form $n+\d$ with $n\in {\bf Z},\ 
\d\in {\bf R}$. The same argument holds here and we 
conclude that the gauge and local Lorentz invariant 
combination $A$ cannot be determined further. The reason
of course is that 
since  $G\ne d C$ one cannot use the 
standard argument to obtain flux quantization. We 
conclude that for four-cycles not wrapping the circle 
$S^1$ we have standard integer or half-integer flux 
quantization, while if the four-cycle 
wraps the $S^1$ we cannot say anything interesting about the 
flux of $G$. 

This paper is organised as follows. In Section 2, we discuss 
preliminaries: conventions and normalisations, and the 
correctly 
modified, periodically-defined step function $\e(\xe)$ on the 
circle, as well as the above-mentioned occurrence 
of an extra factor of $1/3$ when evaluating integrals 
involving $\d \e \e$. In Section 3, we carefully solve the 
modified Bianchi identity for $G$ and discuss 
gauge and local Lorentz invariance and how the global 
definition of $G$ fixes $b$. We also comment on the 
relation to the perturbative heterotic string. Section 4 
discusses anomaly cancellations and the quadratic 
relation between $b$ and $\l^6/\k^4$ is obtained. Section 5 is 
devoted to flux (non) quantization of $G$ and in Section 6 
we spell out the heterotic anomaly-cancelling terms as 
inherited from the present M-theory approach. We conclude
in Section 7 and an appendix summarises the relevant 
anomaly polynomials [\AGW, \GSW] needed in this paper.

{\bf\chapter{Preliminaries}}

\section{Conventions and normalisations}

\noindent
Since in the next section we want to pin down 
various numerical coefficients, normalisations are
important. Our conventions are as follows. 

\subsection{Normalisations of fields and couplings}

\noindent
The anomaly polynomials $I_{12}(1{\rm -loop})_i$, $X_8$ and 
$I_{4,i}$ are defined in the appendix. In particular we 
have 
$$I_{4,i}= {1\over (4\pi)^2} \left( \tr F_i^2 - {1\over 2}
\tr R^2\right) \quad 
\Rightarrow \quad \int_{\C_4} I_{4,i}= m_i -{1\over 2}
p_i\ ,\qquad m_i,p_i \in {\bf Z},
\eqn\di$$
for any four-cycle $\C_4$. For differential forms,
we use the standard normalisation: 
a $p$-form $A^{(p)}$ is given in terms 
of its completely antisymmetric components as
$$A^{(p)}={1\over p!} A^{(p)}_{\m_1 \ldots \m_p} \,
{\rm d}x^{\m_1} \wedge \ldots \wedge {\rm d}x^{\m_p}
={1\over p!} A^{(p)}_{[\m_1 \ldots \m_p]} \,
{\rm d}x^{\m_1} \wedge \ldots \wedge {\rm d}x^{\m_p},
\eqn\dii$$
and $d= {\rm d}x^\m \partial_\m$ so that in particular 
$F=dA+A^2={1\over 2}
F_{\m\n}\,{\rm d}x^\m\wedge {\rm d}x^\n$ and 
for the three-form field of eleven-dimensional supergravity
$C={1\over 3!}C_{\m\n\r}\,
{\rm d}x^\m\wedge {\rm d}x^\n\wedge {\rm d}x^\r$.

The normalisation of the eleven-dimensional supergravity 
action [\CJS] is\foot
{
Our normalisation is related to the one used by Ho\v rava 
and Witten [\HWII] by $C^{\rm HW}_{\m\n\r}= {1\over 
6\sqrt{2}}C_{\m\n\r}$ and 
$G^{\rm HW}_{\m\n\r\s}= {1\over \sqrt{2}}G_{\m\n\r\s}$. 
}
$$ S_{\rm sugra}= \int {\rm d}^{11}x \ e\ \left[ 
-{1\over 2\k^2}{\cal R} 
-{1\over 96\k^2} G_{\m\n\r\s}G^{\m\n\r\s} \right] 
+S_{\rm top} + \ldots
\eqn\diii$$
where ${\cal R}$ is the curvature scalar and $e$ the square 
root of the determinant of the eleven-dimensional metric.
$\k\equiv \k_{11}$ is the eleven-dimensional 
gravitational coupling. 
The so-called topological term is 
of central importance for us:
$$S_{\rm top}= -{1\over 12\k^2}\int C\wedge G \wedge G ,
\eqn\div$$
where $G=d C + \ldots$ Omitted terms indicated by 
\dots involve the gravitino field. 

Similarly, the action for the ten-dimensional 
gauge fields is
$$S_{\rm gauge}=-{1\over 4\l^2} \int {\rm d}^{10}x \  
\sqrt{ g_{10}}\,  F_{AB}F^{AB} +
\ldots \ ,
\eqn\dv$$
with gauge coupling constant $\lambda$ and $F=dA+A^2$.
On the $S^1/{\bf Z_2}$ fixed planes, $g_{10}$ is the
restriction of the eleven-dimensional metric to the planes.

As usual, we use $\m,\n,\ldots$ for eleven-dimensional 
indices and $A,B,\ldots$ for ten-dimensional ones.

\subsection{Engineering dimensions}

\noindent
It is often useful to have in mind the dimensions of the 
various quantities and couplings. Taking
$[{\rm mass}]=1$ as usual, one has $[\k]=-{9\over 2}$ and
$[\l]=-3$. With these two parameters, a dimensionless 
combination is ${\l^3\over \k^2}$ while $[{\k^2\over
\l^2}]=-3$ and this latter combination may be used 
whenever a dimensionful constant is
needed. Gauge and Lorentz connections have dimension
one and then $[R]=[F]=2$ so that $[I_{4,i}]=4$ and
$[X_8]=8$. We also take $[C]=0$ and $[G]=1$ (for forms, 
by $[C]$ we  really mean $[C_{\m\n\r}]$, etc\dots). 

\section{The orbifold $\orb$}

\noindent
We consider the eleven-dimensional manifold as the product 
of a ten-dimensional manifold ${\rm M}_{10}$ and a circle 
$S^1$. For notational convenience, and to 
emphasize that the coordinate on the circle is an angular 
variable (we are working in the upstairs
approach), we use 
\def\xe{\f} 
$$\xe  \in [-\pi,\pi] \simeq S^1 \ ,
\eqn\do$$
rather than $x^{11}$. 
This definition also means that we work in 
general on a circle with unit radius. We will however
reintroduce an arbitrary radius $r$ when discussing the
small-radius limit, {\it i.e.} the perturbative heterotic 
limit. This will be done by assuming that $\f\in [-\pi r,
\pi r]$. The ${\bf Z}_2$
symmetry acts by $\xe\to -\xe$ so that the two fixed 
planes are copies of ${\rm M}_{10}$ at
$\xe=0\equiv \xe_1$ and $\xe=\pi\equiv \xe_2$. We 
then define the one-forms 
$$\d_1=\d(\xe) \,{\rm d}\xe \,\, , 
\qquad\qquad \d_2=\d(\xe-\pi) \,{\rm d}\xe\,\,.
\eqn\dvi$$
These one-forms are well-defined on the circle $S^1$ 
by definition, i.e. $\d_2=\d(\xe-\pi) \,{\rm d}\xe 
\equiv \d(\xe+\pi) \,{\rm d}\xe $. We further need
zero-forms $\epsilon_i$ such that $d\epsilon_i$ includes
$\delta_i$. Determining a primitive of a $\d$-function 
on a circle requires however a minimum of care:
a simple step 
function $\theta(\f)$  will not do the job since it is 
not periodic.
This simple fact apparently has been 
overlooked in previous discussions of
anomaly inflow through the modified Bianchi identity.\foot
{
The issue of 
perodicity is correctly addressed in ref [\LOW]. This paper 
studies background configurations for the four-form $G$ and 
the same terms linear in $\xe$ as in our $\e(\xe)$ below do 
appear. This paper does not discuss anomaly inflow.
}
In fact there cannot be a function $\theta(\xe)$ on the 
circle, i.e. a periodic function 
$\theta(\xe)=\theta(\xe+2\pi)$, such that 
$\theta'(\xe)=\d(\xe)$. Would it exist, one would
also have\foot
{For the integral of a {\it periodic} function  one can 
of course integrate over {\it any} interval of length 
$2\pi$. However, since the orbifold singularities are 
at 0 and $\pi\equiv -\pi$, in order to avoid 
misinterpretations and have each singularity exactly 
once, it is preferable not to integrate from $-\pi$ 
to $\pi$ but to slightly shift the interval of 
integration. Thus, for example,
$\int_{-\pi}^{\pi} d\phi$ is meant to be
${\rm lim}_{\eta\rightarrow0}\int_{-\pi-\eta}^{\pi-
\eta}d\phi$.
} 
$1=\int_{S^1} \d=\int_{-\pi}^{\pi} {\rm d}\xe\, \d(\xe)=
\int_{-\pi}^{\pi} {\rm d}\xe\, \theta'(\xe) =
\theta(\pi)-\theta(-\pi)=0$. The best one can do is 
to define for $\xe\in [-\pi,\pi]$
$$
\e_1(\xe)={\rm sign}(\xe)-{\xe\over \pi} \ .
\eqn\dvii$$
It is periodic, odd under ${\bf Z}_2$, 
and satisfies $\e_1(\pi)=\e_1(-\pi)=0$. 
Similarly one defines $\e_2(\xe)=\e_1(\xe\pm \pi)$. These 
functions satisfy $\e'_1(\xe)=2\d(\xe)-{1\over\pi}$ 
and $\e'_2(\xe)=2\d(\xe-\pi)-{1\over\pi}$. 
In form notation, these definitions 
can be concisely written as
$$ d \e_i= 2\d_i - {1\over \pi} {\rm d} \xe \,,
\qquad\qquad i=1,2.
\eqn\dviii$$
Later on we will need integrals of products of these 
$\e_i$ over the circle. They are elementary and can be
concisely written as
$$ 
\int_{S^1} {\rm d}\xe\  \e_i \e_j = \pi \left( \d_{ij} 
-{1\over 3}\right), \qquad\qquad i,j=1,2,
\eqn\dix$$
where of course here and in the following $\d_{ij}$ stands 
for the Kronecker symbol. Since the $\e_i(\xe)$ 
are odd functions with respect to the ${\bf Z}_2$, 
one also has 
$$\int_{S^1} {\rm d}\xe\ \e_i =0 \ .
\eqn\dixa$$

Another delicate point concerns integrals of products of 
one $\d$ function and two $\e$ functions, 
which are ${\bf Z}_2$-even distributions.
It is often argued 
that one can replace e.g. $\d_1 \e_1 \e_1$ by $\d_1$ 
since due to the $\d$ only the immediate vicinity of 
$\xe=0$ is important, and then $(\e_1(0))^2 \simeq +1$.
This argument is certainly incorrect. 
An expression like $\d_1 \e_1\e_1$ is ill-defined
and has to be regularised. 
One could as well say that $\e_1(0)=0$, implying
$\d_1 \e_1 \e_1=0$ and this would be equally wrong. 
Any sensible regularisation should preserve the one 
relation we want to hold, namely \dviii, as well as
${\bf Z}_2$-oddness of the $\e_i$.
Then indeed in the vicinity of $\xe=0$ the linear
piece $-\xe/\pi$ in $\e_1$ is unimportant
and $\e_1'(\xe) \simeq 2\d(\xe)$ so that 
$$\d(\xe)\e_1(\xe)\e_1(\xe)\simeq {1\over 2}
\e_1'(\xe)\e_1(\xe)\e_1(\xe)
={1\over 6} {{\rm d}\over {\rm d}\xe} \left( 
\e_1(\xe)\right)^3 
\simeq {1\over 6} {{\rm d}\over {\rm d}\xe} \e_1(\xe) 
\simeq {1\over 3}\d(\xe)
\eqn\dx$$
and an important factor ${1\over 3}$ has appeared\foot
{
This factor of ${1\over 3}$ was noticed in ref. [\CONRAD].
}.
This can be rigorously verified using
a specific regularisation\foot
{
One may e.g. take the regularized $\e^\eta_1(\xe)$ to 
be the continuous function which 
coincides with $\e_1(\xe)$ for
$\xe\notin [-\eta, \eta]$ and  equals  $\left( 
{1\over \eta}-{1\over \pi}\right) \xe$
for $\xe\in [-\eta, \eta]$. The corresponding 
$\d^\eta_1(\xe)$ is defined according to Eq. 
\dviii\ and vanishes everywhere except in 
$[-\eta, \eta]$ where it equals ${1\over 2\eta}$. Then
for any test function $f(\xe)$ one has
$\int \d^\eta_1 \e^\eta_1 \e^\eta_1 f
\sim {1\over 3} \left( 1- {\eta\over \pi}\right)^2 f(0)
\to {1\over 3} f(0)$, $\int \d^\eta_1 \e^\eta_1 
\e^\eta_2 f \sim -{1\over 3}{\eta\over \pi} 
\left( 1- {\eta\over \pi}\right) f(0) \to 0$,
$\int \d^\eta_1 \e^\eta_2 \e^\eta_2 f \sim {1\over 3} 
\left( {\eta\over \pi}\right)^2 f(0)\to 0$.
and similarly with 1 and 2 exchanged. Also, $\d^\eta_1
\e^\eta_1$ is an odd function of $\xe\in [-\pi,\pi]$ 
and $\int \d^\eta_1 \e^\eta_1  f = 0$ as well as 
$\int \d^\eta_1 \e^\eta_2  f =0$, and similarly with 
1 and 2 exchanged.
}.
One finally obtains:
$$
\d_i \e_j \e_k \ \ \to \ \ {1\over 3} (\d_{ji} \d_{ki})\  
\d_i \,,\eqn\dxi$$
where again $\d_{ji}$  and $\d_{ki}$ denote Kronecker 
symbols while $\d_i$ is the Dirac $\d$
one-form defined in \dvi. We need also the following 
relation
$$\d_i \e_j  \ \ \to \ \  0 \ .
\eqn\dxii$$

\section{The modified Bianchi identity for $G$}

\noindent
Ho\v rava and Witten [\HWII] consider 
the eleven-dimensional
supergravity action \diii\ on the product of $M_{10}$
with the orbifold $S^1/{\bf Z}_2$, coupled to 
$E_8$ super-Yang-Mills actions located on the 
two $M_{10}$ fixed planes, at $\phi=0$ and $\phi=\pi$.
The orbifold projection eliminates fields odd under 
${\bf Z}_2$. In particular, invariance of $S_{\rm top}$ 
implies that $C_{ABC}$ is odd and projected out, while 
$C_{AB,11}$ is even and kept.
The ${\bf Z}_2$ projection also breaks one half of 
the thirty-two supersymmetries. Preserving the 
sixteen remaining supersymmetries requires however
a modification of the Bianchi identity of the four-form 
field $G$, involving the Yang-Mills curvatures. 
As usual in string effective actions, anomaly 
cancellation in turn requires the appearance in 
this modification of Lorentz curvatures: this contribution
cannot be derived from supersymmetry of effective 
actions with up to two derivatives. 

Instead of $dG=0$, the modified Bianchi identity is 
postulated to be [\HWII]
$$dG=-{\k^2\over \l^2} \sum_i \d_i \left( 
\tr F_i^2-{1\over 2} \tr R^2\right).
\eqn\dxiii$$
As already mentioned, the classical supersymmetry 
calculation only yields the $\tr F^2$ term.
The $\tr R^2$ term is a higher-order effect.
The factor $\kappa^2/\lambda^2$ can be inferred by
simple dimensional analysis, since $[\delta_i]=1$.

\sectionnumber=0
{\bf\chapter{Solution of the modified Bianchi identity}}

\section{A one-parameter family of solutions}

\noindent
We now proceed to solve the modified Bianchi identity 
\dxiii. The combination of $\tr F_i^2$ and $\tr R^2$
that appears is usually called $I_{4,i}$. 
Our normalisations for all relevant anomaly-related 
polynomials in $F$ and $R$ are summarised in 
Appendix A. In particular, we have $I_{4,i}={1\over 
(4\pi)^2} \left( \tr F_i^2 -{1\over 2} \tr R^2\right)$
and thus the Bianchi identity reads
$$dG= - (4\pi)^2\, {\k^2\over \l^2} \sum_i \d_i 
\wedge I_{4,i} \ .
\eqn\ti$$
Since $\d_i$ has support only on the $i^{\rm th}$ 
fixed plane and is a one-form $\sim {\rm d}\xe$, only
the values of the smooth four-form $I_{4,i}$ on this
fixed plane matter and only the components not
including ${\rm d}\xe$ are relevant. For the gauge
piece $\tr F_i^2$ this is automatic, but for the
$\tr R^2$ piece this is a non-trivial statement. It
will prove convenient to use two-dimensional descent
equations $I_4=d \o_3$, $\d \o_3=d \o_2^1$ with all
forms having no ${\rm d}\xe$ component. We will use
the following convention: a tilde on a $p$-form
$A_i^{(p)}$ means that $\til A_i^{(p)}$ is obtained
from $A_i^{(p)}$ by dropping all components $\sim
{\rm d}\xe$ and by setting the argument $\xe=\xe_i$,
i.e. equal to 0 or $\pi$ depending on whether $i=1$
or 2. Clearly, $d$ takes a tilde $p$-form to a tilde
$(p+1)$-form: $d \til A_i^{(p)} =\til {d  A_i^{(p)}}$.
In the Bianchi identity we can then replace $I_{4,i}$
by $\ii$ which has the effect of replacing $R\to
\til R_i \equiv \til R\vert_{\xe=\xe_i}$. One has
$\til R_i= d \til\O_i+\til\O_i\wedge
\til\O_i$ with $ \til\O_i$ the
correspondingly mutilated spin-connection.
Using this $\til\O_i$, one defines the
Chern-Simons three-forms (except for $I_4$, for
convenience of notation, we will no longer write
the degree of the forms explicitly: the subscript
$i=1,2$ refers to the fixed plane, not the degree
of the form)
$$\til \o_i = {1\over (4\pi)^2} \left( \tr ( A_i d A_i +{2\over 3} 
A_i^3) 
- {1\over 2} \tr (  \til\O_i d  \til\O_i +{2\over 3}  
\til\O_i^3) \right)
\eqn\tii$$
which satisfies $d \til\o_i = \ii$. Also under a gauge 
and local Lorentz transformation with parameters 
$\Lambda^{\rm g}$ and $\Lambda^{\rm L}$ 
independent of $\xe$ one has $\d \til\o_i = d \til\o_i^1$ 
where the two-forms $\til\o_i^1$ are
$$\til \o_i^1 = {1\over (4\pi)^2} \left( \tr 
\Lambda^{\rm g} d A_i
- {1\over 2} \tr \Lambda^{\rm L} d \til\O_i\right) \ .
\eqn\tiii$$
Then \ti\ is rewritten as
$$dG=\g \sum_i \d_i \wedge \ii \,, \qquad \ii = 
d \til\o_i \,,\qquad \d \til\o_i = d \til \o_i^1,
\eqn\tiv$$
where to simplify notations we have introduced the 
dimensionful quantity\foot
{
Note that $\g$ can be related to the membrane tension
$T_2$ by $\g=-{2\pi\over T_2}$. We will {\it not} use this 
fact here. We could of course write $T_2$ instead of 
$\g$ in our formulae, but we prefer not to in 
order to emphasize that the discussions and anomaly
cancellations of this paper have nothing to do with membranes
(except for a short comment in Sect. 5).}
$$\g= - (4\pi)^2\, {\k^2\over \l^2} \ .
\eqn\tv$$
This is solved by
$$\eqalign{
G&= d \left( C +{b\over 2} \g \sum_i \e_i\, \oi \right) -
\g \sum_i \d_i \wedge \oi \cr
&= d C + (b-1)   \g \sum_i \d_i \wedge \oi + {b\over 2}
\g \sum_i \e_i\,  \ii
-{b\over 2\pi} \g\,  {\rm d}\xe \wedge \sum_i \oi \,,
}
\eqn\tvi$$
where we used \dviii.
In order to maintain full generality we should allow for
a different parameter $b$ for each fixed plane, i.e.
$G=d \left( C +{1\over 2} \g \sum_i b_i\,  \e_i\,  \oi 
\right) - \g \sum_i \d_i \wedge \oi$. But it is 
clear that the conditions of anomaly cancellation are 
the same for both planes and in the end $b_1$ and $b_2$ 
are determined by the same equation so that $b_1=b_2\equiv b$.
\foot
{
Actually, this equation will turn out to be quadratic 
in $b_i$ and hence has another solution: 
$b_1=-b_2\equiv \pm b$. However, we will see soon that $b_1=+b_2=1$ 
is needed as soon as topologically non-trivial configurations exist.
}
Note the presence of the last term in the second 
expression for $G$. This 
term is a direct consequence of enforcing periodicity 
of the $\e_i(\xe)$ and will turn out to play a most crucial role.

\subsection{Adding five-branes}

We will also consider situations where there are five-branes. 
Each five-brane has a world-volume $W_{6,a}$ where $a$ labels 
the different five-branes. Since five-branes wrapping the 
circle $S^1$ can easily be seen to play no particular role 
in the discussions of section 3.3 and 4.4 below, we will 
concentrate on five-branes that are perpendicular to $S^1$, 
and we denote by $\xe_a$ the coordinate at which they 
intersect the circle. Let there be $N_5$ such five-branes. 
To each of them one associates a brane current 
$\d^{(5)}(W_{6,a})$, analogous to the $\d_i$ of the ten-planes. 
This brane current then contains a piece 
$\d(\xe-\xe_a) {\rm d}\xe$ with the remaining piece 
called $\d^{(4)}(W_{6,a})$. The Bianchi identity for $G$ 
then gets an extra contribution [\WITTENFIVE]
$${\rm d} G\vert_{\rm 5-brane\ contribution}
=\g\sum_{a=1}^{N_5} \d^{(5)}(W_{6,a})
=\g\sum_{a=1}^{N_5} \d(\xe-\xe_a)\, {\rm d}\xe
\wedge \d^{(4)}(W_{6,a}) \ .
\eqn\dgf$$
Accordingly, $G$ gets an extra contribution. Just as before 
when integrating $\d_i \ii$ one now has a choice when 
integrating $\d(\xe-\xe_a) {\rm d}\xe\wedge \d^{(4)}(W_{6,a})$, 
introducing more free parameters:
$$ \eqalign{
G\vert_{\rm 5-brane\ contribution}
=\g\sum_{a=1}^{N_5} \Bigg\{
& {\b\over 2}\left[ \e_1(\xe-\xe_a)\, \d^{(4)}(W_{6,a})
-{1\over \pi}\, {\rm d}\xe\wedge (\theta \d)^{(3)}(W_{6,a}) \right] \cr
& - (1-\b)\, \d(\xe-\xe_a)\,  {\rm d}\xe\wedge (\theta \d)^{(3)}(W_{6,a})
\Bigg\} \ . \cr
}
\eqn\gf$$
It is straightforward to verify, using eqs \dviii\ and \dvi\ 
that the exterior derivative of the r.h.s. of \gf\ indeed 
yields the r.h.s. of \dgf\ for any choice of parameter $\b$ provided
$${\rm d} (\theta \d)^{(3)}(W_{6,a}) =  \d^{(4)}(W_{6,a}) \ .
\eqn\delf$$
Again, there are many different choices for 
$(\theta \d)^{(3)}(W_{6,a})$ but since $\d^{(4)}(W_{6,a})$ 
does not involve the circle coordinate $\xe$  which primitive  
$(\theta \d)^{(3)}(W_{6,a})$ is chosen is irrelevant here.

\section{Gauge and local Lorentz invariance}

\noindent
Now we want $G$ to be gauge and local Lorentz invariant.
While $\ii$ is invariant, the Chern-Simons three-form $\oi$ 
is not and hence $C$ cannot be invariant either. Since
the variation of $C$ will be crucial for anomaly
cancellation, care must be taken. 
The most general $\d C$ giving an invariant 
$G$ is\foot
{
Note that the five-brane contribution to $G$ is 
field-independent and thus does not change $\d C$.
}
$$\d C= d B_2^1 - \g \sum_i 
\left( {b\over 2}\,  \e_i\,  d \oiu + \d_i \wedge 
\oiu\right)
\eqn\tvii$$
with some  two-form  $B_2^1$ linear 
in $\Lambda^{\rm g}$ or $\Lambda^{\rm L}$.  
To determine  $B_2^1$ we must consider the 
${\bf Z}_2$-orbifold projection on the field $C$. 
Recall that $C_{ABC}$ 
is odd and hence projected out, while $C_{AB,11}$ is 
even and kept. But $C_{ABC}=0$ only makes sense if it 
is a gauge invariant statement, i.e. we must have 
$\d C_{ABC}=0$. This yields
$$
(d B_2^1)_{ABC}=\g {b\over 2} \sum_i \e_i\,
(d \oiu)_{ABC}=\g {b\over 2} \Bigl(d\sum_i \e_i\,  
\oiu\Bigr)_{ABC}
$$ 
which is solved by 
$(B_2^1)_{AB}=\g {b\over 2} \sum_i \e_i\,
( \oiu)_{AB}$. 
Hence we choose
$B_2^1=\g {b\over 2} \sum_i \e_i\oiu$ and
$$\eqalign{
\d C&= \g\sum_i \left( {b\over 2}\,   
d\e_i -\d_i\right) \wedge \oiu \cr
&= \g (b-1) \sum_i \d_i\wedge \oiu 
- \g\,   {b\over 2\pi}\,   {\rm d}\xe \wedge 
\sum_i \oiu \ , \cr
}
\eqn\tviii$$
together with
$$C_{ABC}=0 \ .
\eqn\tix$$
Note that $C$ would be gauge invariant in the bulk
(i.e. away from the fixed planes) were it not for the
last term in \tviii, again due to enforcing
periodicity of the ``step functions" $\e_i(\xe)$.
Let us repeat that $\d C$ is such that $\d G=0$.

\section{Global definition of $G$: invariance under large transformations}

So far we have considered invariance under small
transformations that can be continuously deformed to
the identity. However, $G$ must also be invariant under
large gauge and large local Lorentz transformations.
This will be precisely the case if $G$ is globally
well-defined. There is a simple criterion when this
is true  [\WITTENSTRONG]. If $G$ is globally 
well-defined, i.e. $d G$ is exact, then for
any (closed) five-cycle $\C_5$ one  has 
$\int_{\C_5} dG =0$. Take 
$\C_5=\C_4 \times S^1$ where $\C_4$ is an arbitrary 
four-cycle at fixed value 
of $\xe$ or homologous to such a cycle and rewrite 
$dG$ using the Bianchi identity \tiv\ together with \dgf. 
One then gets 
$$\sum_i \int_{\C_4} \ii \equiv {1\over(4\pi)^2}
 \int_{\C_4} \left( \tr F_1^2 -{1\over 2} \tr 
\til R_1^2\right) 
+ {1\over(4\pi)^2}\int_{C_4} \left( \tr F_2^2 
-{1\over 2} \tr \til R_2^2\right) =-N_5
\eqn\tixa$$
where $N_5$ is the number of five-branes intersecting 
$S^1$ and $\C_4$ at a point.
The analogous cohomology condition is well-known from the 
heterotic string. So $G$ will be invariant also under 
large transformations precisely if this condition holds. 
Of course, since we do want $G$ to be invariant we  
assume this condition henceforth.

However, there is one more important piece of information 
we can obtain from requiring global definition of 
$G$ and using Stoke's theorem. Consider a five-dimensional manifold 
${\cal V}$ which intersects exactly one of the fixed 
planes on a four-cycle. For example, we may take it of the form 
${\cal V}=I\times \C$ where $\C$ is a four-cycle and
$I$ is the interval 
$[\xe_1, \xe_2]$ with $-\pi < \xe_1<0$ and $0<\xe_2<\pi$.
Then $\partial {\cal V}=\C(\xe_2)-\C(\xe_1)$ and Stoke's 
theorem gives
$$\int_{\cal V} d G = \int_{\C(\xe_2)} G -\int_{\C(\xe_1)} G\ .
\eqn\tixb$$
Now, the integral on the l.h.s. is evaluated using the 
modified Bianchi identity. The contribution \tiv,  due to the $\d$-function, 
collapses to an integral of $\iu$ over the four-cycle $\C$ 
on the fixed plane. 
The five-brane contribution \dgf\ to ${\rm d}G$ yields an extra 
term $\g N_5(I)$, where $N_5(I)$ is the number of five-branes 
that intersect the interval $I$ and the four-cycle $\C$.
On the other hand, the integrals on the 
r.h.s are evaluated using the solution \tvi\ and \gf\  for $G$. Only 
the components $G_{ABCD}$ contribute, i.e. from \tvi\ only the piece 
${b\over 2}\g \sum_i \e_i \ii$ and from \gf\ only the piece 
${\b\over 2}\g \sum_a \e_1(\xe-\xe_a) \d^{(4)}(W_{6,a})$ contribute. 
For the integral at 
$\xe=\xe_2$ one has $\e_1(\xe_2)=1-{\xe_2\over \pi}$ and 
$\e_2(\xe_2)=-{\xe_2\over \pi}$, while for the  integral at 
$\xe=\xe_1$ one has $\e_1(\xe_1)=-1-{\xe_1\over \pi}$ and 
$\e_2(\xe_1)=-{\xe_1\over \pi}$. Of course
$\int_{\C(\xe_2)} \ii = \int_{\C(\xe_1)} \ii = \int_{\C} \ii$
since $\ii$ is independent of $\xe$.
Furthermore $\int_{\C(\xe_2)} \e_1(\xe-\xe_a) \d^{(4)}(W_{6,a})
- \int_{\C(\xe_1)} \e_1(\xe-\xe_a) \d^{(4)}(W_{6,a})$ equals
${\xe_1-\xe_2\over \pi}$ if the five-brane does not intersect 
the interval $I$ and equals $2+{\xe_1-\xe_2\over \pi}$ if it 
does intersect $I$. Using \tixa\ and collecting all the pieces, 
eq. \tixb\ becomes
$$\g \int_{\C} \iu + \g N_5(I) 
= {b\over 2} \g \left( 2  \int_{\C}\iu
- {\xe_1-\xe_2\over \pi} N_5 \right) 
+{\b\over 2}\g \left( 2 N_5(I)
+ {\xe_1-\xe_2\over \pi} N_5 \right)
\ .
\eqn\tixc$$
But the exact positions of $\xe_1$ and $\xe_2$ are arbitrary, 
and upon slightly varying them such that $N_5(I)$ remains 
unchanged one concludes that the terms linear in $\xe_1-\xe_2$ 
and the terms independent of  $\xe_1-\xe_2$ must vanish separately. 
This yields two equations:
$$\eqalign{
(\b-b) N_5 &=0 \ , \cr
(1-\b) N_5(I) + (1-b) \int_{\C} \iu  &=0 \ . \cr
}
\eqn\tixd$$
%
%
Had we chosen the five-manifold ${\cal V}$ to intersect the 
other fixed plane we would have obtained an analogous equation
with $\id$. 

Let first $N_5=0$ so that one simply gets 
$(1-b)\, \int_{\C} \iu =0$. There are two possibilities. If firstly 
$\int_{\C}\iu = 0$,
each of the two source terms in the modified 
Bianchi identity is cohomologically trivial, 
so that the 
$\til\o_i$ can be globally well-defined. In this case we see 
that $b$ is unconstrained. Secondly, if
$\int_{\C}\iu \ne 0$  each of the two source 
terms individually is cohomologically non-trivial (although 
their sum is trivial by \tixa), and one must take $b=1$.
Note that $b=1$ eliminates the terms in $G$ containing $\d$-functions 
so that  $G$ becomes finite everywhere on $S^1$. 

Now let $N_5\ne 0$. Then 
the first eq. \tixd\ gives $\b=b$ and the second eq. gives
$(\int_{\C} \iu +N_5(I)) (1-b) =0$. Upon varying the interval 
$I$ we may change $N_5(I)$ (unless all five-branes are stuck on 
the fixed planes) and conclude that $b=1$. Thus global definition 
of $G$ fixes $b=\b=1$. The resulting four-form $G$ then reads:
$$
\eqalign{
G\vert_{b=1,\, \b=1}=d C 
&+{\g\over 2} \left[ \sum_{i=1,2} \e_i\, \ii 
+ \sum_{a=1}^{N_5} \e_1(\xe-\xe_a)\, \d^{(4)}(W_{6,a}) \right]\cr
&-{\g\over 2\pi}\, d\xe \wedge \left[ \sum_{i=1,2} \oi 
+ \sum_{a=1}^{N_5}\, (\theta \d)^{(3)}(W_{6,a}) \right] \ .\cr
}
\eqn\tixe$$

To summarize, with the ${\bf Z}_2$ projection \tix\ on
the three-index tensor, global definition of $G$
corresponds to the transformation \tviii\ and the 
cohomology condition \tixa, as well as \tixd.

\section{The case $b=1$}

\noindent
As just noted, $b=1$ is required whenever $\int \ii \ne 0$
or five-branes not wrapping the circle are present.
In this case  delta-function
singularities are absent from $G$.
The case $b=1$ also presents some other interesting and 
important 
features when considering Eqs. \tvi\ and \tviii\ from the 
point of view of reduction to the perturbative heterotic string.
For this we take $N_5=0$.

Expanding $G_{ABC,11}$ and $C_{AB,11}$ in Fourier modes 
along $S^1$ straightforwardly leads to
$$
\eqalign{
G^{(0)}_{ABC,11} &= d_{[A} C^{(0)}_{BC],11}
 +{\gamma\over2\pi}[\til\o_1+\til\o_2]_{ABC}, \cr
G^{(n)}_{ABC,11} &= d_{[A} C^{(n)}_{BC],11} 
-{\gamma\over2\pi}(b-1)
[\til\o_1+(-1)^n\til\o_2]_{ABC}, \qquad\qquad n>0.
}
\eqn\expa
$$
The zero mode is $b$-independent.
Since $G$ is gauge and local Lorentz invariant we see that 
$C^{(0)}_{AB,11}$ can be neither gauge nor local 
Lorentz invariant. 
On the other hand, the higher modes of $C_{AB,11}$ 
are gauge and local Lorentz invariant if and only if $b=1$.
To make contact with ten-dimensional
heterotic strings in the field-theory (perturbative) 
limit, we want to truncate $C_{AB,11}$ to its zero-mode
only, and this truncation is  safe only if the higher modes
are gauge invariant. This again points towards $b=1$. 

In the next section we will show that anomaly cancellations 
relate $b^2$ and $\l^6/\k^4$, but do not fix one or the other.
However, the gauge and local
Lorentz variation of the topological term will have a 
$b$-dependent contribution which is a variation of a
(local) counterterm and hence does not contribute to the
formal twelve-form which characterizes the anomaly. 
The appearance of this term is related to the $b$-dependent
contributions in the variation of $C$, and then 
to the higher modes in its expansion. 
On the other hand, a direct calculation of the 
anomaly-cancelling terms, by first truncating $C$ to
$C^{(0)}$ and then calculating the resulting ten-dimensional
action leads directly to a Green-Schwarz term 
which corresponds to $b=1$, the case in which all 
truncated modes are gauge 
invariant. This will be done in Sect. 6.

The conclusion  then is that  anomaly cancellation alone  
does not fix $b$. The perturbative heterotic
limit however (the small $S^1$ radius limit) selects $b=1$ 
because it ensures gauge invariance of the massive modes. 
This condition is  essentially due to compactification 
on a small space. As we have seen in the previous subsection,
global considerations also impose $b=1$, provided 
topologically non-trivial configurations occur. 

\section{A consistency check: reduction to the heterotic string}

\noindent
When the circle $S^1$ is very small the perturbative 
heterotic string theory must emerge. The fields of the 
latter are defined (modulo possible rescalings) as the 
zero-modes of the Fourier expansion on the circle of the 
corresponding M-theory fields. In particular, 
as mentioned already
$$\eqalign{
B_{AB} \equiv C^{(0)}_{AB,11} 
&= {1\over2\pi r}\int_{-\pi r}^{\pi r} C_{AB,11}(\xe)\, 
{\rm d}\xe \ , \cr
H_{ABC} \equiv G^{(0)}_{ABC,11} 
&= {1\over2\pi r}\int_{-\pi r}^{\pi r} G_{ABC,11}(\xe)\, 
{\rm d}\xe \ ,\cr
}
\eqn\tixb$$
where we have introduced an arbitrary radius
$r$  of $S^1$. 
Integrating  the $ABC,11$ components of the solution 
\tvi\ for $G$ over the circle $S^1$ yields
$$H= dB+{\g\over2\pi r}
\sum_i \til\o_i  \ .
\eqn\tixc$$
As already observed in \expa,
the parameter $b$ disappears thanks to the last term in \tvi,
a term due to the condition of $S^1$ periodicity 
of the ${\bf Z}_2$-odd ``step function" $\e_i(\xe)$. 
In the small radius limit, the Lorentz Chern-Simons 
forms contained in $\til\o_1$ and $\til\o_2$ become 
the same and we can replace
$\til\o_1+\til\o_2$ by\foot
{
$ \O_{3{\rm YM}}$ and $\O_{3{\rm L}}$ are conventionally 
normalised Chern-Simons three-forms with 
$d \O_{3{\rm YM}}=\tr F_1^2+\tr F_2^2$ and
$d \O_{3{\rm L}}=\tr R^2$. Of course, they should not be 
confused with the spin connection one-forms 
$\til\O_1, \til\O_2$ and $\til\O$ used some time ago.
} 
${1\over (4\pi)^2} 
\left( \O_{3{\rm YM}}-\O_{3{\rm L}} \right)$ 
and \tixc\ becomes
$$H=dB - {\k^2\over 2\pi r\l^2}
\left(  \O_{3{\rm YM}}-\O_{3{\rm L}}  \right)  \ .
\eqn\tixd$$
Upon reexpressing $\k\equiv \k_{11}$ in terms of 
the ten-dimensional gravitational coupling 
$\k_{10}$ and rescaling the fields to one's favorite 
normalisations, one obtains the standard relation for 
the heterotic string. In our case, 
${\kappa^2\over2\pi r} = \kappa^2_{10}$, and we redefine 
$H$ and $B$ according to
$$
H = {\kappa_{10}^2\over\lambda^2} \wh H, \qquad\quad
B = {\kappa_{10}^2\over\lambda^2} \wh B,
\eqn\hetdef
$$
so that one obtains the standard relation of the 
heterotic string:
$$
\wh H = d\wh B -  \O_{3{\rm YM}} + \O_{3{\rm L}} \ .
\eqn\tixdd
$$
The gauge variation of $\wh B$ can be directly obtained
from its definition and the expression of $\delta C$,
Eq. \tviii. The result is as expected
$$
\delta\wh B = \O_{2{\rm YM}}^1 - \O_{2{\rm L}}^1,
$$
with $\delta\O_{3{\rm YM}} = d\O_{2{\rm YM}}^1$\
and $\delta\O_{3{\rm L}}=d\O_{2{\rm L}}^1$. Of course,
$\wh H$ is invariant.

It is easy to see how these relations are modified by 
the presence of five-branes: an M-theory five-brane not 
wrapping $S^1$ (with $\b=1$) only contributes a zero-mode 
${\g\over 2\pi} (\theta \d)^{(3)}(W_{6,a})$ to 
$G_{ABC,11}$. Such a term then appears on the r.h.s. 
of \tixdd\ and $d \wh H$ gets an extra term 
$\sim \d^{(4)}(W_{6,a})$ which is the appropriate 
source term from a heterotic NS five-brane.

To complete the comparison with the low-energy effective 
supergravity of the heterotic string, we compute the Einstein
and Yang-Mills terms and the kinetic terms for $\wh H$ using 
the basic Lagrangians \diii\ and \dv. With the metric
ansatz
$$
g_{\mu\nu} = \pmatrix{ \varphi^{-1/4}g_{AB} & 0 \cr
0 & -\varphi^2 },
$$
we obtain, in terms of ten-dimensional quantities only, 
$$
{\cal L}_{\rm het} 
= -{\sqrt{g_{10}}\over2\kappa_{10}^2}
\Bigl[ {\cal R} 
+{1\over2}{\kappa_{10}^2\over\lambda^2}\varphi^{-3/4}
\tr(F_{AB}F^{AB})
-{1\over12} 
\left({\kappa_{10}^2\over\lambda^2}\varphi^{-3/4}\right)^2
\wh H_{ABC}\wh H^{ABC}\Bigr] 
+\ldots
\eqn\hetlagr
$$
Notice that $G_{ABCD}$ would contribute to four-derivative 
terms only. Finally, since $\lambda^2\kappa_{10}^{-3/2}$ is
dimensionless, it can be absorbed in a redefinition of the
field $\varphi$ to obtain
$$
{\cal L}_{\rm het} 
= -{\sqrt{g_{10}}\over2\kappa_{10}^2}\Bigl[
{\cal R} +{1\over2}\kappa_{10}^{1/2}\varphi^{-3/4}
\tr (F_{AB}F^{AB})
-{1\over12} \kappa_{10}\,\varphi^{-3/2}
\wh H_{ABC}\wh H^{ABC}\Bigr]
+\ldots,
\eqn\hetlagrb
$$
and the field-dependent gauge coupling constant is as usual
$$
g^2 = \kappa_{10}^{3/2} \varphi^{3/4}.
$$
This shows that ${\cal L}_{\rm het}$ has two parameters, 
$\kappa_{10}$ and the expectation value of $\varphi$ 
(related to the radius $r$ of the circle), and that
$\lambda$ cannot be observed in the heterotic limit. 

\sectionnumber=0
{\bf\chapter{Anomaly cancellations}}

\noindent 
The ${\bf Z}_2$ orbifold projection generates a chiral 
spectrum and, as a consequence, a chiral gauge, 
mixed and gravitational quantum anomaly in $M_{10}$ 
which can be characterized by a formal twelve-form. 
We have seen that the modification of the Bianchi 
identity is at the origin of a well-defined gauge 
and Lorentz variation of the three-form field $C$. 
This variation will in turn produce an anomalous 
variation of the action. We now study the sources 
for this anomaly inflow and prove anomaly cancellation. 
To begin with, we will not consider five-branes. 
They will be taken into account separately in sect. 4.4.
In this section we still keep $b$ as a parameter, since 
from the M-theoretic point of view, the requirement $b=1$ 
occurs in the first place only if $\int \ii \ne 0$. 
Even in the topological 
sector where $\int \ii = 0$ there are anomalies which 
need to be cancelled. In any case we will find that 
anomaly cancellation relates $b^2$ to $\l^6/\k^4$. 
In the sector where $\int \ii \ne 0$ and $b=1$ this 
then fixes the ratio $\l^6/\k^4$. But this ratio 
should not depend on the topological sector of the 
theory and this in turn will imply that $b=1$ always.

\section{Anomaly inflow from $S_{\rm top}$}

\noindent
It is now straightforward to determine the anomaly 
inflow due to the topological term $S_{\rm top}$:
$$\d S_{\rm top}= - {1\over 12 \k^2} \int \d C \wedge 
G\wedge G \ .
\eqn\tx$$
Since $\d C$ only contains components with 
a ${\rm d}\xe$, only the components of $G$ not 
containing ${\rm d}\xe$ can contribute, i.e. 
$G_{ABCD}$. Remembering that $C_{ABC}=0$, we get from 
\tvi\ $G_{ABCD}= {b\over 2} \g \sum_j \e_j 
(\ij)_{ABCD}$ and hence 
$$\d S_{\rm top}= - {\g^3 b^2\over 48 \k^2} \int \sum_i 
\left[ (b-1) \d_i- {b\over 2\pi} {\rm d} \xe \right] 
\oiu\, \sum_j \e_j \ij \, \sum_k \e_k \ik \ .
\eqn\txi$$
Now we use \dix\ and \dxi\ to perform the ${\rm d}\xe$ 
integrals over $S^1$. In the integrals {\it not} 
involving $\d_i$ it is important that, apart from 
the $\e_j \, \e_k$, the rest of the integrand, 
namely $\oiu \ij \ik$ is independent of $\xe$. 
The purpose of introducing the tilde quantities 
was precisely to make manifest this $\xe$-independence. 
We get
$$\eqalign{
\d S_{\rm top} &= - {\g^3 b^2\over 48 \k^2} \left[
{1\over 3} (b-1) \int_{M_{10}} \sum_i \oiu \ii\ii 
-{b\over 2} \int_{M_{10}} \sum_{i,j,k} (\d_{jk}  
-{1\over 3}) \oiu \ij \ik \right] \cr
&\equiv \d S_{\rm top}^{(1)} + \d S_{\rm top}^{(2)} \ .
}
\eqn\txii$$
The second term $\d S_{\rm top}^{(2)}$ looks quite 
different from the first one $\d S_{\rm top}^{(1)}$,
especially due to the multiple sum over $i,j,k$. 
However, we will see that both terms correspond to
the {\it same} anomaly polynomial $I_{12}$. Carrying
out the sums in $\d S_{\rm top}^{(2)}$, one gets
$$
\d S_{\rm top}^{(2)}={\g^3 b^3\over 144 \k^2} 
\int_{M_{10}} (\til\o^1_1 + \til\o^1_2)  \left( (\iu )^2 + 
(\id )^2 -\iu\id\right) \ .
\eqn\txiii$$
Upon applying the descent equations, the corresponding 
invariant formal twelve-form is
$$\eqalign{
I_{12}^{({\rm top},2)} &={\g^3 b^3\over 144 \k^2} (\iu+\id)
 \left( (\iu )^2 + (\id )^2 -\iu\id\right) \cr
&= {\g^3 b^3\over 144 \k^2} \left( (\iu )^3
+(\id )^3\right).
}
\eqn\txiv$$
But this is of the {\it same} form as the
invariant twelve-form corresponding to 
$\d S_{\rm top}^{(1)}$:
$$
I_{12}^{({\rm top},1)} =-{\g^3 b^2 (b-1)\over 144 \k^2}  
\left( (\iu )^3+(\id )^3\right) ,
\eqn\txv$$
and adding them up, we see that the terms cubic in $b$
exactly {\it cancel} so that the twelve-form 
characterizing anomaly inflow from the topological 
term reads
$$
I_{12}^{({\rm top})} = + {\g^3 b^2 \over 144 \k^2}  
\sum_i (\ii )^3 
= - {\pi \over 3} \left( { (4\pi)^5\over 12} 
{\k^4\over \l^6} b^2 \right)
\sum_i (\ii )^3 \ .
\eqn\txvi$$
Since the field $G$ is real, the parameter $b$
is real as well and $b^2$ is necessarily positive, as
is ${\k^4\over \l^6}$. So the sign of the coefficient
in $I_{12}^{({\rm top})}$ is fixed. This reflects the 
fact that $N=1$ ten-dimensional supersymmetry is chiral
and specific signs appear in the Bianchi identity once
the gravitino chirality is chosen. In any case, the 
above sign is as required to cancel the quantum 
anomaly generated by chiral fermions.  

It is quite amazing that the two so different 
looking anomalies 
$\d S_{\rm top}^{(1)}$ and $\d S_{\rm top}^{(2)}$ 
correspond to the same anomaly twelve-form (apart from
the different $b$-dependence). However, using the 
descent equations, as explained in Appendix A, 
it is not difficult to explicitly find a local 
ten-dimensional counterterm that relates the two 
forms:
$$
\int_{M_{10}} \sum_{i=1,2} \oiu\, (\ii )^2 
= \int_{M_{10}} (\til\o^1_1 + \til\o^1_2)  
\left( (\iu )^2 + (\id )^2 -\iu\id\right) 
+ \d \, \int_{M_{10}} \Delta_{10}
\eqn\txvii$$
with
$$\eqalign{
\Delta_{10} 
&= {2\over 3}\  \left( \til\o_1+\til\o_2\right)
\left( \til\o_1\iu +\til\o_2\id 
-{1\over 2} \til\o_1\id  -{1\over 2} \til\o_2 \iu 
\right) \cr
&=\til\o_1\, \til\o_2\, (\id-\iu) \ . \cr
}
\eqn\txviia$$
One may then rewrite $\d S_{\rm top}$ as
$$\d S_{\rm top}={\g^3 b^2\over 144 \k^2} \left[ \int_{M_{10}} 
\left( \til\o^1_1 (\iu)^2 + \til\o^1_2 (\id)^2 \right)
- b \d \int_{M_{10}} \Delta_{10} \right] \ .
\eqn\txviib$$
This form will be useful in Sect. 6.

\section{Anomaly inflow from the Green-Schwarz term}

\noindent
We still have to determine the variation of the 
Green-Schwarz term $\int G\wedge X_7$. Here arises 
the question whether one should take $\int G\wedge 
X_7$ or $\int C\wedge X_8$ with $X_8=d X_7$ and 
$X_8={1\over (4\pi)^3\, 12}\left({1\over 2}\tr R^4 
- {1\over 8}(\tr R^2)^2\right)$ (cf. Eq. (A.5) in 
the Appendix). Note that $X_8$ obeys $d X_8=0$ and $\d X_8=0$. 
Both choices are equivalent if $G=dC$, but 
differ at present. Going through  the same steps 
with $\int C\wedge X_8$ as above with $\int C\wedge 
G\wedge G$ we will show below that this form does 
not allow to cancel the one-loop anomalies. The 
correct form is $\int G\wedge X_7$. The appropriate 
normalisation of this term is known independently 
[\VAFAWITTEN, \DUFF], but we will rederive it 
from the present anomaly cancellation. Let
$$S_{\rm GS} ={c\over \g}\int G\wedge X_7
\eqn\txviii$$
where $\g=-(4\pi)^2 {\k^2\over \l^2}$ as before 
and $c$ is a dimensionless constant to be 
determined. One has (using descent equations 
$X_8=d X_7$, $\d X_7=d X_6^1$)
$$\eqalign{
\d S_{\rm GS} 
&={c\over \g}\int G\wedge \d X_7
={c\over \g}\int G\wedge d X_6^1
= - {c\over \g}\int d G\wedge X_6^1 \cr
&= -c \int \sum_i \d_i\wedge \ii \wedge X_6^1 
= - c \int_{M_{10}} \sum_i \ii \wedge 
\til X_{6,i}^1 \ . 
}
\eqn\txix$$
Note the replacement of the bulk $X_6^1$ by the 
$\til X_{6,i}^1$ defined only on the $i^{\rm th}$ 
plane. Since $\ii$ is closed and gauge/local 
Lorentz invariant, this corresponds to the twelve-form
$$I_{12}^{({\rm GS})} = -c \sum_i \ii \wedge \til X_{8,i} \ .
\eqn\txx$$

Had we started with $\wh S_{\rm GS} ={c\over \g} 
\int C\wedge X_8$, we would have obtained, 
using Eq. \tviii,
$$ \d \wh S_{\rm GS}   ={c\over \g} \int \d C\wedge X_8
=c\, (b-1) \int_{M_{10}} \sum_i \oiu\wedge \til X_{8,i} 
- c\, {b\over 2\pi} \int \sum_i {\rm d}\xe 
\wedge \oiu\wedge X_8 \ .
\eqn\txxa$$
While in the first term the $\d$-function has the 
effect of replacing $X_8$ by $\til X_{8,i}$, in 
the second term $X_8$ truly depends on $\xe$. We see 
that the first term alone corresponds to a polynomial
$\wh I_{12}^{({\rm GS},1)} =c (b-1)  \sum_i \ii 
\wedge \til X_{8,i} $ 
and has the right form. Its coefficient however is wrong:
if $b=1$ it vanishes and if one is allowed to take a $b\ne 1$ 
one would be forced into a non-standard choice of $c$. 
In any case, in the second term 
however, $X_8$ genuinely depends on the circle 
coordinate $\xe$ and there is no way to make it 
equal to $ \til X_{8,i} $ on the $i^{\rm th}$ plane 
as needed for anomaly cancellation. We conclude that 
$\wh S_{\rm GS}$ could be suitable at best only for $b=0$. 
But this case is certainly ruled out as it would lead to 
$\d S_{\rm top}=0$. Note that the possibility to 
discriminate between $S_{\rm GS}$ and $\wh S_{\rm GS}$ 
relies on the presence of the second term in \txxa\ 
and this term again is a consequence of enforcing 
periodicity of the ``step functions" $\e_i(\xe)$.

\section{Cancellation of the one-loop anomaly}

\noindent
The sum $I_{12}^{({\rm top})} + I_{12}^{({\rm GS})}$ 
must cancel the one-loop anomaly given in the 
Appendix, Eq. (A.5). (These expressions refer to 
ten-dimensional anomalies on a given fixed plane 
and should be understood as involving only 
tilde quantities.) It is
$$I_{12}(1{\rm -loop}) = \sum_i \left( {\pi\over 3} (\ii )^3
 + \ii \wedge \til X_{8,i} \right) \ .
\eqn\txxi$$
We get anomaly cancellation, 
$I_{12}^{({\rm top})} + I_{12}^{({\rm GS})} 
+ I_{12}(1{\rm -loop}) =0$, 
if and only if
$$ b^2={12\over (4\pi)^5} {\l^6\over \k^4}
\eqn\txxii$$
and
$$ c=1 \ .
\eqn\txxiii$$
While $c=1$ was known previously, from cancellation
of the anomaly due to a five-brane (see below), 
earlier literature claims a cubic equation for the 
parameter $b$. Instead, we have shown here that a careful
treatment of periodicity along $S^1$ cancels the cubic
terms and leads to a quadratic relation.

Using eqs. \txviia, \txix\ and \txxii\ we may now rewrite 
the total anomalous variation as
$$
\d S_{\rm top}+\delta S_{\rm GS} 
= -{\pi\over3}\int_{M_{10}}
\left[ \til\o_1^1 \iu^2 
+ \til\o_2^1 {\id}^2 
- b\, \d \Delta_{10}\right]
-\int_{M_{10}} \left[ \iu\til X_{6,1}^1
+ \id\til X_{6,2}^1\right] \ .
\eqn\totalvariation
$$
Note that the term $\sim b$ is the variation of a local 
ten-dimensional counterterm, and this is why it
does not contribute to the twelve-form of the descent equations. 
We will see in Section 6 that in the small
$S^1$ radius limit corresponding to the perturbative heterotic
string, this particular form of the anomalous variation
is at the origin of a local counterterm usually unexpected from
standard ten-dimensional arguments. 

\section{The five-brane anomaly}

\noindent
The same Green-Schwarz term \txviii\ is also 
able to cancel any additional five-brane anomaly\foot
{
For five-branes, there is also the normal bundle 
anomaly, but it can be taken care of independently [\NM] 
and we will not consider it here.
}.
The latter is a purely gravitational anomaly of the 
six-dimensional chiral theory living on the 
worldvolume of the five-brane (a chiral tensor 
multiplet). The  invariant eight-form corresponding
to this one-loop anomaly   simply is
$$I_8^{\rm 5-brane}(1{\rm -loop}) = X_8 \,,
\eqn\txxiv$$
where the sign on the r.h.s actually depends 
on the choice of chirality. The 
Bianchi identity for $G$ is further modified by the 
presence of the five-brane as given by \dgf.
Recall that $W_6$ denotes the five-brane worldvolume
and $\delta^{(5)}(W_6)$ is a five-form such that for any 
six-form $I_6$, 
$\int_{M_{11}} \,\delta^{(5)}(W_6)\wedge I_6 = \int_{W_6}\,I_6$.
(A possible switch of chirality would produce a 
minus sign on the r.h.s. of \txxiv\ and \dgf.) 
Then one sees immediately from \txix\ that there is 
an additional five-brane contribution to  
$\d S_{\rm GS}$:
$$\d S_{\rm GS}\vert_{\rm 5-brane}
= - {1\over \g} \int d G\vert_{\rm 5-brane} \wedge X_6^1
= - \int \d^{(5)}(W_6) \wedge X_6^1 = - \int_{W_6} X_6^1,
\eqn\txxvi$$
which is the descendent of $I_8^{({\rm GS})}=-X_8$ thus 
cancelling the $I_8^{5-{\rm brane}}(1{\rm -loop})$.

We should still check that the new five-brane contribution 
to $dG$ does not spoil our previous results.
This is a crucial point which, to the best of our knowledge, 
has not been addressed at the level of M-theory
anomaly cancelling terms.\foot
{
A discussion of anomaly cancellation in a 
Calabi-Yau background with
five-branes can be found in [\LOWCY].
} 
We will 
see that the result is quite non-trivial. Indeed, 
when solving the Bianchi identity, $G$ has an 
extra piece:
$$G\vert_{\rm  5-brane} = G\vert_{\rm without\ 
five-brane} 
+ \g (\theta\d)^{(4)}(W_6)
\eqn\txxvii$$
where $d  (\theta\d)^{(4)}(W_6) = \d^{(5)}(W_6)$. 
In section 3.1, we have explicitly given the form of 
$ (\theta\d)^{(4)}(W_6)$ for five-branes not wrapping the circle,
and later on in sect 3.3 we have shown that in this case the 
parameter $\b$ must be taken to be 1. For five-branes that wrap 
the circle, $\d^{(5)}(W_6)$ does not contain $\xe$ and any 
primitive $(\theta\d)^{(4)}(W_6)$ is allowed so far.
For the time being we do not specify which type of five-branes 
we consider and thus we keep the more generic notation 
$(\theta\d)^{(4)}(W_6)$.
What is important is 
that the extra piece in $G$ does not depend on any 
fields and is thus trivially gauge and local Lorentz 
invariant. Thus the variation $\d C$ as given in 
\tviii\ is unchanged. The variation $\d S_{\rm GS}$ is modified by 
the additional piece \txxvi\ as required. What about 
$\d S_{\rm top}=-{1\over 12 \k^2} \d \int C\wedge 
G\wedge G\ $? 
Since  $\d C$ is the same as before, 
the only additional contribution is due to the 
additional piece in $G$. We then get an additional 
piece in $\d S_{\rm top}$:
$$\eqalign{
\d S_{\rm top}\vert_{\rm  5-brane}
&= -{1\over 6 \k^2} 
\int \d C \wedge G\vert_{\rm without\ 5-brane} \wedge 
G\vert_{\rm extra} \cr
&= -{\g^3 b\over 12 \k^2}\int  
\sum_i \left[ (b-1) \d_i -{b\over 2\pi} {\rm d}\xe 
\right] \wedge \oiu \wedge \sum_j \e_j \ij \wedge  
(\theta\d)^{(4)}(W_6) \,,
}
\eqn\txxviii$$
where we used \tviii\ for $\d C$. 

If the five-brane wraps the circle, $\d^{(5)}(W_6)$ and 
$(\theta\d)^{(4)}(W_6)$ are independent of $\xe$ and 
then the only $\xe$-dependence of the above integrand is 
$ \left[ (b-1) \d_i -{b\over 2\pi} {\rm d}\xe \right] \e_j $
which integrates to zero by virtue of Eqs. \dixa\ and \dxii.
Hence the anomalous variation of $S_{\rm top}$ is 
not affected by these five-branes.

If the five-brane worldvolume does not extend in 
the $\xe$-direction, things are more subtle. 
We have seen that in this case necessarily $b=1$ and $\b=1$.
Then using the explicit form of \gf\ with $\b=1$ or eq. \tixe,
eq. \txxviii\ becomes
$$
\d S_{\rm top}\vert_{\rm  5-brane}
= {\g^3 \over 48\pi \k^2}\int  
\sum_i {\rm d}\xe 
 \wedge \oiu \wedge \sum_j \e_j \ij \wedge  
\sum_{a=1}^{N_5} \e_1(\xe-\xe_a) \d^{(4)}(W_{6,a}) 
\eqn\txxviii$$
where we have now explicitly summed over all such five-branes.
The integral over the circle can be performed explicitly, as well 
as the four integrals involving the $\d^{(4)}(W_{6,a})$. The 
result is a non-vanishing contribution localised on the 
world volumes of the five-branes:
$$
\d S_{\rm top}\vert_{\rm  5-brane}
= {\g^3 \over 48\pi \k^2} \sum_{a=1}^{N_5}\ \int_{W_{6,a}} 
(\til\o_1^1 + \til\o_2^1) \left( f_1(\xe_a) \iu
 + f_2(\xe_a) \id \right) \ ,
\eqn\txxix$$
where the $f_i(\xe_a)= \int {\rm d}\xe\,  \e_i(\xe) \e_1(\xe-\xe_a)$ 
are given by 
$f_1(\xe)={2\pi\over 3} - 2\vert\xe\vert +{\xe^2\over \pi}$ and 
$f_2(\xe)=-{\pi\over 3} +{\xe^2\over \pi}$ and obey 
$f_i'(\xe)=-2\e_i(\xe)$. Note that they  are ${\bf Z}_2$ 
even functions. Thus we see that there is an extra anomaly 
inflow from $S_{\rm top}$ into the six-dimensional theory on 
the five-brane world volumes.
The associated invariant eight-form is
$$I_8^{({\rm top,\ 5-brane})} 
= {\g^3 \over 48\pi \k^2} \sum_{a=1}^{N_5}\ 
\left[ f_1(\xe_a) \left( (\iu)^2+\iu \id \right)
+  f_2(\xe_a) \left( (\id)^2 +\iu \id \right) \right]
\eqn\txxixa$$
which clearly is non-vanishing.
This is an interesting new effect.
A somewhat related issue appears in [\LOWCY] where 
it is found, within the context of  Calabi-Yau compactifications 
with less supersymmetry, 
that there are gauge fields that originate on the five-branes.
The consequences of the non-vanishing term \txxix\ go beyond the scope
of the present paper
and will be discussed  elsewhere [\BDS].

\sectionnumber=0
{\bf\chapter{The issue of $G$-flux  quantization}}

\section{Does flux quantization hold?}

\noindent
Cancellation of all anomalies only required the 
validity of relation \txxii\ between $b^2$ and $\l^6/\k^4$. 
To put it differently, whatever the ratio $\l^6/\k^4$ 
is, there is a choice of the parameter $b$ that 
cancels all anomalies. On the other hand, 
we have seen that if there are topologically non-trivial 
gauge/gravity configurations such that $\int_{\C_4} \ii \ne 0$, 
one is forced to take $b=1$ in order to have a globally 
well-defined four-form field $G$. We will now explore the 
consequences of this fact for the flux of $G$ and compare 
with Witten's result on flux quantization [\WITTENFLUX]
obtained in the downstairs approach 
which in a certain way also corresponds to a fixed 
value of $b$. 

We will evaluate the integral of $G$ on the two different 
categories of four-cycles. To simplify the discussion, we suppose
that there are no five-branes. Modifications due to the 
latter can be trivially incorporated.

\subsection{Four-cycles not wrapping $S^1$}

\noindent
Consider a  four-cycle $\C(\xe)$ 
not wrapping the $S^1$ and at a fixed value of 
$\xe\in (0,\pi)$. Then
$${2\over \g}  \int_{\C(\xe)} G  
= b \sum_i \e_i(\xe) \int_{\C(\xe)} \ii
= b \left[ \left( 1-{\xe\over \pi}\right) \int_{\C(\xe)} \iu
-{\xe\over \pi} \int_{\C(\xe)} \id \right] 
= b \int_{\C(\xe)} \iu 
\eqn\qv$$
where the $\xe$-dependent terms have cancelled thanks to the 
cohomology condition \tixa. As shown in section 3.3, 
either $\int_{\C(\xe)} \iu=0$ and $b$ is arbitrary, or 
$\int_{\C(\xe)} \iu = n_1-{1\over 2} p_1\ne 0$ with 
$n_1, p_1 \in {\bf Z}$ in which case $b=1$. In either case 
we get
$${2\over \g} \int_{\C} G =   \left( n_1-{1\over 2} 
p_1\right) 
= -   \left( n_2-{1\over 2} p_2\right) 
\quad \hbox{for any 4-cycle } \C \hbox{ not wrapping } S^1 \ .
\eqn\qix$$
Usually one mentions another contribution to this 
integral which would come from the $dC$ piece in $G$. 
But with $C_{ABC}=0$ there is no such contribution here. 

This looks much like Witten's flux quantization [\WITTENFLUX] 
which was obtained in the downstairs approach and which reads 
${1\over \g_{\rm downstairs}} \int_{\C} G 
= \left( n_1-{1\over 2} p_1\right)$. Although the two 
conditions seem to differ by a factor of two, they are 
actually the same. The point is that the gravitational 
couplings $\k^2$ in the upstairs and downstairs approach 
differ precisely by this factor of two [\CONRAD]: 
$\k^2_{\rm downstairs}=\k^2_{\rm upstairs}/2$ and since 
$\g=-(4\pi)^2 \k^2/\l^3$ one also has 
$\g_{\rm downstairs}=\g_{\rm upstairs}/2$. Of course, 
in all our formulas $\g\equiv \g_{\rm upstairs}$ and one 
sees that eq \qix\ is exactly Witten's flux quantization. 
Using Eq. \txxii\ with $b=1$ one can  reexpress 
this flux quantization with only $\l$ or only $\k$ appearing 
as coefficient, but this is not too illuminating.

\subsection{Four-cycles wrapping $S^1$}

\noindent
We must also consider four-cycles of the 
form $\C=\C_3\times S^1$ where $\C_3$ is some 
three-cycle at fixed value of $\xe$. 
Obviously, such cycles do not exist in the downstairs approach 
and we cannot expect the corresponding flux to be related 
to Witten's quantization condition. An arbitrary
four-cycle then is homologous to $m_1$ times this 
four-cycle $\C$ plus $m_2$ times any of the four-cycles 
of the previous subsection. One has from Eq. \tvi\
$$
\int_{\C_3\times S^1} G= \int_{\C_3\times S^1} 
\left( dC +  \g\sum_i \left[ (b-1) \d_i -{b\over 2\pi}  
{\rm d}\xe \right]
\wedge \til\o_i  \right)
=  \int_{\C_3\times S^1}  dC - \g \int_{\C_3} 
\sum_i  \til\o_i \ .
\eqn\qx$$
As always when integrating over $S^1$, 
the $b$-dependent terms have cancelled. This integral 
is closely related to the flux $\int H = \int 
( d B - \o_{\rm YM} + \o_{\rm L})$ in the heterotic 
string. Indeed, as discussed in Section 3.5, it 
exactly reduces to the latter in the small radius limit. 
The flux of $H$ was discussed some time ago [\ROHMWITTEN] 
and it was concluded (when appropriately normalised) 
to be of the form $n+\d$ with $n\in {\bf Z},\ \d\in 
{\bf R}$. A similar argument holds here and we conclude 
that the value of \qx\  cannot be determined further. 
The key difference with Witten's analysis [\WITTENFLUX] 
is that since $G\ne d C$ one cannot use the standard 
argument to obtain flux quantization. 

We conclude that 
for four-cycles not wrapping the circle $S^1$ we have 
the standard flux quantization, 
while if the four-cycle wraps the $S^1$ we cannot say 
anything interesting about the flux of $G$. 

\section{A remark on the membrane action}

\noindent
Let us first review the standard argument [\WITTENFLUX]. 
In uncompactified M-theory or M-theory on a smooth 
manifold one simply has $G=dC$. Since this needs not 
hold globally one typically argues that $\int_{\C}G= 
\int_{\C^+} d C^{(+)} + \int_{\C^-} d C^{(-)} = 
\int_{\C_3} ( C^{(+)}  - C^{(-)} )$ where $\C_3=\partial 
\C^+=-\partial \C^-$. 
But since the membrane action $T_2 \int_{\C_3} C + 
\ldots$ ($T_2$ is the membrane tension)
should be well-defined modulo $2\pi$ one 
concludes that $T_2 \int_{\C}G = 2\pi n$ with integer 
$n$. Witten has argued [\WITTENFLUX] that the 
three-dimensional membrane theory does in certain cases have a 
so-called parity anomaly [\AGDPM] which is a sign 
ambiguity $e^{i\pi p}, \ p\in {\bf Z}$ of the fermion 
determinant. This implies that actually the 
well-definedness of the membrane functional integral 
requires $T_2 \int_{\C}G = 2\pi(n-{1\over 2}p)$. This 
fits with the flux quantization since $T_2=-{2\pi\over \g}$.
In our present upstairs discussion however, this  
relation for the $G$-flux holds  
(with the appropriate redefinition of 
$\k^2$ by a factor of two as discussed above) if $\C$ 
does not wrap the $S^1$, but is replaced by \qx\ if it does. 
Does this mean that the membrane functional integral is 
no longer well-defined? Of course not. First, the above 
argument is spoiled since $G\ne dC$ everywhere. Second, 
a coupling of $C$ to the membrane worldvolume $\C_3$ of 
the type $\int_{\C_3} C $ without modification certainly 
does not lead to a well-defined functional integral 
since we have observed that $C$ is neither gauge nor 
local Lorentz invariant. Obviously then there must be 
corrective terms to restore the invariance. It would 
be interesting to explicitly construct these terms. 
A clue is probably provided by Eq. \qx\ which is gauge 
and local Lorentz invariant.

{\bf\chapter{The heterotic anomaly-cancelling terms}}

\noindent
As a last check of our scheme, we compute in the
small-radius limit the anomaly-cancelling terms of the
heterotic string. Taking the zero modes along the 
circle amounts to identifying $\til R_1$ and $\til R_2$ with 
$R$ as well as the seven-form 
$X_7$ in $M_{11}$ with its restriction 
$\til X_{7,1}\sim \til X_{7,2}\sim\til X_7$ to $M_{10}$
(no ${\rm d}\xe$ components) and similarly for $X_8=d X_7$ 
and $\til X_{8,1}\sim \til X_{8,2}\sim\til X_8$. We 
recall that $X_{8,i}$ is given in the appendix.
As before, $B_{AB}$ is identified with the zero mode
${1\over2\pi r}\int_{S^1}C_{AB,11}$, cf. Eq. \tixb, and 
$C_{ABC}=0$ as usual.
Using Eq. \tvi\ (with an obvious insertion of a factor 
of ${1\over r}$ in its last term), 
the Green-Schwarz term $\int G\wedge X_7$ 
then reduces to
$$\eqalign{
{1\over\gamma}\int G\wedge X_7 \quad&\to\quad
{2\pi r\over\gamma}\int_{M_{10}}B \wedge \til X_8 
+\int_{M_{10}}\int_{S^1}\sum_i\left[(b-1)\delta_i\wedge\til\o_i
-{b\over2\pi r}d\phi\wedge\til\o_i\right]\wedge \til X_7
\cr
&= -{2\pi r\over 12 (4\pi)^5}{\lambda^2\over\kappa^2}
\int_{M_{10}}B \wedge 
\Bigl[ {1\over2}\,\tr  R^4 -{1\over8}(\tr  R^2)^2 \Bigr] 
-\int_{M_{10}}\sum_i \til\o_i\wedge\til X_7\,.
}
\eqn\hetgsa
$$
Similarly, the topological term yields, using Eq. \dix\ to 
perform the integral over $S^1$ and Eq. \txxii\ to reexpress 
$b^2$ in terms of $\l^6/\k^4$,
$$\eqalign{
-{1\over12\kappa^2}\int C\wedge G\wedge G \quad&\to\quad
-{b^2\gamma^2\over 48\kappa^2}
\int_{M_{10}}\int_{S^1} C\wedge
\Bigl(\sum_i\epsilon_i\til I_{4,i}\Bigr) \wedge
\Bigl(\sum_j\epsilon_j\til I_{4,j}\Bigr)
\cr
&\to -{2\pi r b^2\gamma^2\over 144\kappa^2}\int_{M_{10}} B\wedge
\Bigl((\iu)^2 + (\id)^2 
-\iu\id\Bigr) \cr
&= -{2\pi r\over 12 (4\pi)^5}{\lambda^2\over\kappa^2} 
\int_{M_{10}} B\wedge
\Bigl[ (\tr  F_1^2)^2 + (\tr  F_2^2)^2
-(\tr  F_1^2)(\tr  F_2^2) \cr
&\hskip2.4cm
+{1\over4}(\tr  R^2)^2
-{1\over2}(\tr  R^2)(\tr  F_1^2 +\tr  F_2^2)\Bigr]\,.
}
\eqn\hetgsb
$$

At this point it is useful to switch to differently normalised 
heterotic  variables: the ten-dimensional gravitational 
constant is $\kappa_{10}^2= \kappa^2/(2\pi r)$ and $B$ 
is rescaled to  the heterotic $\wh B$ as in Eq. \hetdef. 
It is also convenient 
to define\foot
{
To facilitate comparison with the literature, we note e.g. 
that the $X_8$ used in the textbook by Green, Schwarz and 
Witten [\GSW] is ${1\over 4} \wh X_8$.
}
the heterotic eight-form $\wh X_8$ and the seven-form
$\wh X_7$ such that $d \wh X_7=\wh X_8$:
$$
\eqalign{
\wh X_8 =&\,\, (\tr  F_1^2)^2 + (\tr  F_2^2)^2
-\tr  F_1^2 \, \tr  F_2^2
-{1\over2}\tr  R^2 \, (\tr F_1^2 +\tr F_2^2) \cr
&\,\,
+{1\over2}\,\tr R^4 +{1\over8}(\tr  R^2)^2 \ ,
\cr
\wh X_7 =&\,\, \O_{3,1}\, \tr F_1^2
+ \O_{3,2}\, \tr F_2^2
-{1\over2}\O_{3,1}\, \tr F_2^2
-{1\over2}\O_{3,2}\, \tr F_1^2 \cr
&\,\,
-{1\over4}\O_{3,{\rm L}}(\tr F_1^2 +\tr F_2^2) 
-{1\over4}(\O_{3,1}+\O_{3,2})\, \tr R^2
+{1\over2}\O_{7{\rm L}} +{1\over8}\O_{3{\rm L}}\, \tr R^2\ . 
}
\eqn\hetgse
$$
In the last expression, the gauge and gravitational 
Chern-Simons three-forms $\O_{3,1}$, $\O_{3,2}$ and 
$\O_{3{\rm L}}$ and seven-form
$\O_{7{\rm L}}$ are 
such that $d\O_{3,i}=\tr F_i^2$, $i=1,2$ and
$d\O_{3{\rm L}}=\tr R^2$, $d\O_{7{\rm L}} = 
\tr R^4$ so that indeed $\wh X_7$ verifies $d\wh X_7=\wh X_8$. 
Note that our previous definitions $\til X_7$ and 
$\til X_8$ only had 
gravitational contributions, while now, following the 
standard heterotic notation, $\wh X_7$ and $\wh X_8$ have 
both, gravitational and gauge, as well as mixed contributions.
With these definitions, the sum of the contributions \hetgsa\ and
\hetgsb\ becomes
$$
S_{\rm GS, het} 
= -{1\over 12 (4\pi)^5}\int_{M_{10}}\, 
\left( \wh B\wedge \wh X_8\ 
+\  (\O_{3,1}+\O_{3,2}-\O_{3{\rm L}})\wedge 
({1\over 2}\O_{7{\rm L}} - {1\over 8}\O_{3{\rm L}}\tr R^2 )
\right) 
\eqn\siv$$
which we may rewrite as
$$
\eqalign{
S_{\rm GS, het} 
=& -{1\over 12 (4\pi)^5}\int_{M_{10}}\, 
\left( \wh B\wedge \wh X_8 \ 
+\  (\O_{3{\rm YM}}-\O_{3{\rm L}})\wedge \wh X_7 \right) \cr
& +{1\over 8 (4\pi)^5}\int_{M_{10}}\,
(\O_{3,1}-{1\over 2}\O_{3{\rm L}})
(\O_{3,2}-{1\over 2}\O_{3{\rm L}})
(\tr F_2^2-\tr F_1^2) \ .
}
\eqn\sv$$
The first line is the standard 
Green-Schwarz counterterm of the heterotic string [\GSW], 
while the
second term is a local counterterm specific to the 
form of the anomaly-cancelling term generated by M-theory 
on $S^1/{\bf Z}_2$. Of course, it is our old friend from 
Eqs. \txviia\ and \txviib, namely
${\pi\over 2} \int_{M_{10}}\, \Delta_{10}
={3\over 2}\left(-{\g^3 b^3\over 144\k^2}
\int_{M_{10}}\Delta_{10}\right)$
with $b=1$. 
As already explained in Sect. 3.4, the small-radius 
limit automatically produces this term with a coefficient 
corresponding to $b=1$.
The extra factor of ${3\over 2}$ with respect to Eq. \txviib\
is due to the usual choice of including only part of this 
term, namely $-{1\over 2}$ of it, into the standard heterotic 
counterterm of the first line of \sv, leaving us with a 
factor of $+{1\over 2} + 1 ={3\over 2}$
for the second term.

{\bf\chapter{Conclusions}}

\noindent
We have carefully studied the solution for $G$ of the modified 
Bianchi identity. It appeared to depend on an arbitrary 
parameter $b$. When requiring $G$ to be globally well-defined, 
i.e. invariant under small and large gauge and local Lorentz 
transformations we have discovered that this parameter $b$ is 
fixed to $b=1$ in all topologically non-trivial sectors of the 
theory (i.e. with non-vanishing gauge/gravitational instanton 
numbers and/or with five-branes)
Anomaly cancellation equates 
$b^2$ and ${12\over (4\pi)^5} \, {\l^6\over\k^4}$, independently
of the choice of topological sector.
In those sectors where $b=1$ this fixes 
${\l^6\over\k^4}={(4\pi)^5\over 12}$. Since this ratio cannot 
depend on the topological sector of the theory, anomaly 
cancellation in turn  implies that $b=1$ always.

Thus anomaly cancellation and global well-definedness of $G$ 
have selected exactly one solution $G$ to the modified Bianchi 
identity. This is {\it neither a consequence} of flux 
quantization {\it nor} has it anything to do with membrane 
actions. Instead, we observe that for four-cycles not wrapping 
$S^1$, the flux of $G$ indeed automatically obeys Witten's flux 
quantization while for cycles wrapping the $S^1$ the flux is 
more general. Also, in this $\orb$ orbifold for membranes 
wrapping the $S^1$ the naive membrane action is not gauge/local 
Lorentz invariant and needs to be modified.

We  observed that $b=1$ is also very natural when 
considering the reduction to the heterotic string in the small 
radius limit, since it is only for $b=1$ that all higher modes 
of the Fourier expansion on the circle can be consistently 
decoupled. We explicitly wrote out the anomaly cancelling terms 
obtained from this reduction. They differ from what is usually 
taken in the heterotic string by the addition of a well-defined 
local counterterm.

Anomaly cancellations in the presence of five-branes 
are more subtle: For  five-branes 
not wrapping the circle there is an extra contribution from 
$S_{\rm top}$  signalling that non-trivial things are happening 
on those five-branes.

It is quite amazing that this simple $\orb$ orbifold of M-theory 
still had that many subtle points to be revealed. Now that we 
understand them, it is relatively straightforward to study 
other more complicated orbifolds of M-theory. This will be 
done elsewhere [\BDS].


{\bf\Appendix{A}}

\def\A{{\cal A}}
\noindent
In this appendix, we give the relevant anomaly polynomials used 
in the paper. Different authors have conventions differing by 
factors of $2\pi$, etc\dots We use the conventions of [\GSW], 
which were also used in the nice appendix of [\FLO]. 
Recall that an anomaly $\A_n$ in $n$ dimensions, i.e. $\d S = 
\int d^n x\, \A_n$, is described by a unique 
gauge-invariant\foot
{
By ``gauge" invariant, we really mean invariant under gauge 
and local Lorentz transformations. Similarly, the variation 
$\d$ designates both types of infinitesimal transformations.
}
polynomial $I_{n+2}$ in $n+2$ dimensions (a twelve-form
in our ten-dimensional case)
which is related to $\A_n$ by the descent equations
$$I_{n+2}=d I_{n+1}\ , \qquad\qquad \d I_{n+1} = d \A_n\ .
\eqn\api$$
Recall that this is so because the anomaly $\A_n$ itself is 
not uniquely defined. Indeed we always have the freedom 
to add an $n$-dimensional local counterterm to the action 
$S$: $S\to S'=S+\int d^n x\, \D_n$, so that $\A_n\to 
\A'_n=\A_n+\d\D_n+d\D_{n-1}$, with an arbitrary $\D_{n-1}$. 
On the other hand, the descent equations imply that 
$I_{n+1}$ is defined up to a $d\D_n$. But then
$\d I_{n+1}$ is defined up to a $\d d\D_n= d \d \D_n$ 
meaning that $\A_n$ is only defined up to a $\d \D_n$ and 
a $d \D_{n-1}$ as it should. When one applies the descent 
equations to different ways of writing the {\it same} 
invariant anomaly polynomial $I_{n+2}$ one is naturally 
led to different anomalies $\A_n$ and $\A'_n$, the two 
differing by the gauge variation $\d \D_n$ of a local 
counterterm and possibly a total derivative $d\D_{n-1}$.

The relevant contributions to the gauge, gravitational 
and mixed anomalies due to the different chiral fields in 
ten dimensions are given by the following invariant 
twelve-forms of the gauge and Lorentz curvatures:
$$\eqalign{
I^{(3/2)}_{\rm grav}(R)&= {1\over (2\pi)^5 \, 6!} 
\left( {55\over 56} \tr R^6 
- {75\over 128} \tr R^4\,  \tr R^2 +{35\over 512} 
(\tr R^2)^3\right), \cr
I^{(1/2)}_{\rm grav}(R)&= {1\over (2\pi)^5 \, 6!} 
\left( -{1\over 504} \tr R^6 - {1\over 384} \tr R^4\, 
\tr R^2 -{5\over 4608} (\tr R^2)^3\right), \cr
I^{5{\rm -form}}_{\rm grav}(R)&= {1\over (2\pi)^5 \, 6!} 
\left( -{496\over 504} \tr R^6 + {7\over 12} \tr R^4\, 
\tr R^2 -{5\over 72} (\tr R^2)^3\right), \cr
I^{(1/2)}_{\rm mixed}(R,F)&= {1\over (2\pi)^5 \, 6!} 
\left( {1\over 16} \tr R^4\,   \Tr F^2 
+ {5\over 64} (\tr R^2)^2\,   \Tr F^2 -{5\over 8} 
\tr R^2\, \Tr F^4\right), \cr
I^{(1/2)}_{\rm gauge}(F)&= {1\over (2\pi)^5 \, 6!}\  
\Tr F^6 , \cr
}
\eqn\apii$$
where $\Tr$ denotes as usual the trace over generators
of the adjoint representation of the gauge group: the 
gauge curvature two-form $F$ has values in  the
adjoint representation. An additional
factor 1/2 must be included for Majorana-Weyl fermions, and 
opposite chiralities contribute with opposite signs. 

The one-loop anomaly generated by each of the $S^1/{\bf Z}_2$
fixed planes includes an $E_8$ gauge (and mixed) contribution,
and a gravitational contribution that is the half of what 
would be expected in ten dimensions. This is because
of the coupling of eleven-dimensional supergravity to 
ten-dimensional fields [\HWI]. A pragmatic way to 
understand this extra factor of one half is to realise that, 
in the limit where the two fixed planes coincide, 
the sum of the anomalies of the two fixed planes should 
reproduce the standard ten-dimensional $E_8\times E_8$ 
anomaly. Hence the one-loop anomaly on the $i^{\rm th}$
plane ($i=1,2$) for a general gauge group is
$$\eqalign{
I_{12}(1{\rm -loop})_i&={1\over 2}\times {1\over 2} 
\left( I^{(3/2)}_{\rm grav}(R) - I^{(1/2)}_{\rm grav}(R)\right)\cr
&+ {1\over 2} \left( n_i\, I^{(1/2)}_{\rm grav}(R) + 
I^{(1/2)}_{\rm mixed}(R,F_i)
+ I^{(1/2)}_{\rm gauge}(F_i) \right),\cr
}
\eqn\apiii$$
where the overall factor of ${1\over2}$ is due to the 
Majorana-Weyl condition and the extra
factor of ${1\over2}$ in  the gravitational term has just 
been explained. The dimension of the gauge group on the 
$i^{\rm th}$ plane is denoted by $n_i$.
Note that $R$ is meant to be located on the $i^{\rm th}$ 
plane. In the following, we write $R_i$ to emphasize this fact. 
Such an anomaly has a chance to be cancelled by an appropriate
Green-Schwarz mechanism only if this twelve-form factorises 
into a four-form and an eight-form. For
this to work the $\tr R^6$ term must vanish and the 
$\Tr F^6$ term must be expressible
entirely as a combination of $\Tr F^4\, \Tr F^2$ and 
$(\Tr F^2)^3$. The first condition selects $n_i=248$ while 
the only appropriate factorisation of $\Tr F^6$ is obtained
if
$$ \Tr F_i^6= {1\over7200}(\Tr F_i^2)^3, \qquad\qquad
\Tr F_i^4 = {1\over100}(\Tr F_i^2)^2\,.
\eqn\apiv$$
These two conditions only hold for $E_8$, which also has the
required dimension $n_i=248$. The presence of large integers
in the coefficients is due to the large value 30 of the 
quadratic Dynkin index for the adjoint representation of 
$E_8$. Expressions become nicer if one redefines the trace 
as $\Tr=30\, \tr$. One then arrives at the factorised form
$$\eqalign{
I_{12}(1{\rm -loop})_i&= {\pi\over 3} (I_{4,i})^3 + I_{4,i} 
X_{8,i} \ , \cr
I_{4,i}&= {1\over (4\pi)^2} \left( \tr F_i^2 - {1\over 2} 
\tr R_i^2\right) \ , \cr
X_{8,i}&={1\over (4\pi)^3\,  12} \left( {1\over 2} 
\tr R_i^4 - {1\over 8} (\tr R_i^2)^2\right) \ . \cr
}
\eqn\apv$$

One sees that each of the $I_{12}(1{\rm -loop})_i,\  i=1,2$ 
is factorised. For the perturbative $E_8\times E_8$ heterotic 
string the relevant polynomial is the sum 
$I_{12}(1{\rm -loop})_1 + I_{12}(1{\rm -loop})_2$. How can 
this also be  factorised? For the perturbative limit 
(weak coupling) the two fixed planes coincide and one should 
take the same value for $R$ in both expressions. Then the 
purely gravitational $X_8$ is the same and one gets
$$\eqalign{
I_{12}(1{\rm -loop})_1 + I_{12}(1{\rm -loop})_2 
&\longrightarrow (I_{4,1}+I_{4,2}) X_8
+{\pi\over 3} \Bigl ( (I_{4,1})^3+ (I_{4,2})^3\Bigr)\cr
&= (I_{4,1}+I_{4,2}) \left[ X_8 +{\pi\over 3}\Bigl( 
(I_{4,1})^2 + (I_{4,2})^2
-I_{4,1} I_{4,2} \Bigr) \right], \cr
}
\eqn\apvi$$
which is factorised thanks to the trivial identity $a^3 + b^3 
=(a+b) (a^2+b^2-ab)$. It is now easy to rewrite this 
expression in term of traces over the representations
of the product group $E_8\times E_8$ to recover the 
well-known expression of the anomaly for
$E_8\times E_8$ super Yang-Mills coupled to $N=1$ 
supergravity (heterotic string). But we do not need 
it here.

We have chosen the normalisation of the four-forms $I_{4,i}$ 
such that their integrals over a four-cycle equal the
{\it integral} characteristic class of the $E_8$ bundle 
minus a quarter of the {\it even} first Pontryagin class. 
More precisely, for any four-cycle $\C_4$ one has
$$\int_{\C_4} I_{4,i} = m_i - {1\over 2} p_i \ , \quad m_i,p_i \in 
{\bf Z} \ .
\eqn\apvii$$

\vskip9mm

\ack
J.-P. D. has benefitted from discussions with D. 
L\"ust, H. P. Nilles and B. A. Ovrut. 
This work was partially supported by the Swiss National Science 
Foundation, by the European Union under TMR contract 
ERBFMRX-CT96-0045 and by the Swiss Office for Education and 
Science. 

\vskip6mm

\refout

\end